\documentclass{aa}  

\usepackage{hyperref}
\usepackage{footnote} 
\usepackage{graphicx}
\usepackage{amsmath}
\usepackage{mathabx}
\usepackage{txfonts}
\usepackage{xcolor}
\usepackage[graphicx]{realboxes}
\usepackage{enumerate}
\usepackage[toc,page]{appendix}

\begin{document}

\title{Gas disk interactions, tides and relativistic effects in the rocky planet formation at the substellar mass limit}

  \subtitle{}

  \author{Mariana B. S\'anchez
          \inst{1,2}\thanks{e-mail:msanchez@fcaglp.unlp.edu.ar}
          \and
          Gonzalo C. de El\'ia\inst{1,2}
          \and
          Juan Jos\'e Downes \inst{3}
          }
  \institute{Facultad de Ciencias Astron\'omicas y Geof\'isicas, Universidad Nacional de La Plata, Paseo del Bosque S/N (1900), La Plata, Argentina.\\
            \and
             Instituto de Astrof\'isica de La Plata, CCT La Plata-CONICET-UNLP, Paseo del Bosque S/N (1900), La Plata, Argentina.\\
             \and
             Departamento de Astronom\'ia, Facultad de Ciencias, Universidad de la Rep\'ublica, Igu\'a 4225, 14000, Montevideo, Uruguay.\\
             }
  \date{}

 
\abstract
{The confirmed exoplanet population around very low mass stars is 
increasing considerable through data from the latest space missions and 
improvements in ground-based observations, particularly with the 
detection of Earth-like planets in the habitable zones. However, 
theoretical models need to improve in the study of planet formation 
and evolution around low-mass hosts.}
{Our main goal is to study the formation of rocky planets and the
first $100~\textrm{Myr}$ of their dynamical evolution around 
a star with a mass of $0.08 M_\odot$, which is close to the 
substellar mass limit.}
{We developed two sets of $N$-body simulations assuming an 
embryo population affected by tidal and general relativistic effects, refined by the inclusion 
of the spin-up and contraction of the central 
star. This population is immersed in a gas disk during the first 10 Myr.
Each set of simulations incorporated a different 
prescription from the literature to calculate the interaction between the gas-disk and the embryos: one widely used prescription which is based on results from hydrodynamics simulations, and a recent prescription that is based on the analytic treatment of dynamical friction.}
{We found that in a standard disk model, the dynamical evolution 
and the final architectures of the resulting rocky planets are strongly 
related with the prescription used to treat the interaction within the 
gas and the embryos. Its impact on the resulting close-in planet
population and particularly on those planets that are located inside the habitable zone is particularly strong.}
{The distribution of the period ratio of adjacent confirmed exoplanets  
observed around very low mass stars and brown dwarfs and the exoplanets that we
obtained from our simulations agrees well only when the prescription based
on dynamical friction
for gas-embryo interaction was used. Our results also reproduce a close-in planet population of interest that is located inside the habitable zone. A fraction of these planets will be exposed for a long period of time to the stellar irradiation inside the inner edge of the evolving habitable zone until the zone reaches them.}

\titlerunning{Rocky planet formation at the substellar mass limit}
  \maketitle
%

\section{Introduction}

During the past decade, the search of planets around 
very low mass stars (VLMSs) has increased significantly even with the 
confirmation of Earth-like planets around brown dwarfs (BDs) through 
transit and radial velocity observations mainly from the 
\textit{Keppler/K2} missions, the \textit{HARPS} (High Accuracy Radial velocity Planet Searcher) and \textit{CARMENES} (Calar Alto high-Resolution search for M dwarfs with Exoearths with Near-infrared and optical Échelle Spectrographs) spectrographs, and 
the \textit{Trappist} (Transiting Planets and Planetesimals Small Telescope) telescope
\citep{Muirhead2012,Gillon2017,Astudillo2017,Grimm2018,Crossfield2019,Zech2019,Dreizler2020,Sabotta2021}. The close-in rocky planets that are hosted by such low-mass objects inside the habitable zones are ideal targets for the search of life in the solar neighborhood. Therefore, future space missions such as \textit{PLATO} (Planetary Transits and Oscillations of stars) and \textit{JWST} (James Webb Space Telescope) will be able to detect and atmospherically characterize Earth-like planets in the habitable zones around M dwarfs \citep[][]{Rauer2014}. 
From a theoretical perspective, only few models have been
developed to study the rocky planet formation around VLMSs.
Initially \citet[][]{Payne2007,Raymond2007,Ciesla2015,Liu2020} predicted
planetary systems with more compact orbits than the systems that revolve around more
massive stars. Moreover, they found that the mass of the planets increases with the mass of the host. 
Further results showed the relevance of tidal
and general relativistic effects for the orbital changes of 
these compact systems. The contraction
and spin-up of VLMSs during the pre-main sequence phase also allows
the planet population to follow different dynamical paths
\citep[e.g.,][]{Barnes2010,Heller2010,Bolmont2011,Bolmont2013,Sanchez2020}.
Recently \citet[][]{Coleman2019} studied the formation of the
Trappist-1 system by incorporating the migration and orbital 
damping of the planetary embryos that is caused by torques exerted
by the gas disk \citep[e.g.,][]{Papaloizou2000,Tanaka2004,Paardekooper2010,Paardekooer2011,CN08,Ida2020} and showed its relevance in the early dynamic 
evolution of the system.

The standard initial mass function predicts that VLMSs and BDs are the most abundant objects in the solar
neighborhood. The planetary formation around these objects is therefore
essential for estimating the probability of Earth-like 
planets in habitable zones. Their proximity and number makes 
them relevant targets in the search for potentially habitable planets.
\citep[e.p.][]{Kasting1993,Selsis2007,Kopparapu2013,Barnes2013}.

We study the rocky planet formation around a star 
with a mass of 0.08 M$_\odot$, which is close to the substellar mass limit.
We conducted N-body simulations considering an embryo population orbiting
the star. We included tidal and general relativistic effects and the
contraction and spin-up of the star during the first 100 Myr 
of its evolution. We incorporated the gas-disk interactions onto the embryos during the first 10 Myr by assuming a standard disk model
and using two different prescriptions for the corresponding torques: the classic formulas from \citet[][]{CN08}, and the recent results from \citet[][]{Ida2020}.
Our aim is to make a comparative analysis of the two prescriptions 
for the gas-disk torques through a dynamical analysis along the gas and post-gas stages and in the final architectures of the resulting planetary systems. We also compare these systems with observed counterparts.

In Section \ref{sec:diskmodel} we describe the standard disk model. In Section \ref{sec:disk-embryo interaction} 
we explain the implementation of the two prescriptions for the gas 
torques and the set of test simulations we conducted to guarantee the 
agreement between the numerical and analytical models. 
In Section \ref{sec:simulations} we characterize the initial parameters of our N-body simulations. In Section 
\ref{sec:results} we develop a detailed analysis of the planet 
dynamics and architectures during the gas and post-gas stages. 
Finally, Section \ref{sec:conclusiones} summarizes our conclusions.

\section{Disk model}
\label{sec:diskmodel}

We adopted the disk model from \citet{Ida2016} in which the structure is 
described by the gas surface density profile $\Sigma_{\textrm{g}}$, the disk temperature 
 $T_{\textrm{g}}$, and the gas-disk aspect ratio $h_{\textrm{g}}=H_{\textrm{g}}/r$, where $r$ is the radial coordinate in the midplane of the disk, and $H_{\textrm{g}}$ is the gas scale height, which depends on the 
 heating process. The model includes two different dominant heating mechanisms: 
 the internal viscous dissipation for the inner disk and the irradiation from 
 the central star for the outer disk. For the inner region of the disk 
 $\Sigma_{\textrm{g}}$, $T_\textrm{g}$ and $h_\textrm{g}$ profiles are given by:
 
 \begin{equation}
 \begin{split}
 \Sigma_{\textrm{g},\textrm{vis}}= &2100\left(\frac{\dot{M_{\textrm{g}}}}{10^{-8}~\textrm{M}_\odot ~\textrm{yr}^{-1}}\right)^{3/5}\left(\frac{M_{\star}}{\textrm{M}_{\sun}}\right)^{1/5}\left(\frac{\alpha_{\textrm{g}}}{10^{-3}}\right)^{-4/5}
     \\
     &\left(\frac{r}{\textrm{au}}\right)^{-3/5}~\mathrm{g~cm^{-2}}
\end{split}
 \end{equation}

 \begin{equation}
 \begin{split}
 T_{\textrm{g},\textrm{vis}}= &200\left(\frac{\dot{M_{\textrm{g}}}}{10^{-8}~\textrm{M}_\odot ~\textrm{yr}^{-1}}\right)^{2/5}\left(\frac{M_{\star}}{\textrm{M}_{\sun}}\right)^{3/10}\left(\frac{\alpha_{\textrm{g}}}{10^{-3}}\right)^{-1/5}
     \\
     &\left(\frac{r}{\textrm{au}}\right)^{-9/10}~ \mathrm{K}
\end{split}
 \end{equation}

  \begin{equation}
 \begin{split}
 h_{\textrm{g},\textrm{vis}}= &0.027\left(\frac{\dot{M_{\textrm{g}}}}{10^{-8}~ \textrm{M}_\odot ~\textrm{yr}^{-1}}\right)^{1/5}\left(\frac{M_{\star}}{\textrm{M}_{\sun}}\right)^{-7/20}\left(\frac{\alpha_{\textrm{g}}}{10^{-3}}\right)^{-1/10}
     \\
     &\left(\frac{r}{\textrm{au}}\right)^{1/20}  
\end{split}
 \end{equation}
 where $M_\star$ is the mass of the central object, $\dot{M_\textrm{g}}$ 
 is the gas accretion rate and $\alpha_\textrm{g}$ is the viscous coefficient 
 related to the viscosity $\nu=\alpha_\textrm{g} c_\textrm{s}T_\textrm{g} H_\textrm{g}$, 
 where $c_\textrm{s}$ is the sound speed at the temperature of the disk midplane at a 
 given radial distance \citep{Shakura1973}.
 We assumed that the inner disk is optically thick with an average opacity
 $\kappa=1~\mathrm{cm^{2}~g^{-1}}$.

In order to smooth $\Sigma_\textrm{g,vis}$ at the inner edge of the disk we 
multiplied it with the term $\tanh[(r-r_0)/(r_0h_0)]$, where $r_0$ and $h_0$ are
the radius and aspect radius at the inner edge respectively, as was suggested 
by \cite{Cossou2014}, \cite{Matsumura2017}, and \cite{Brasser2018}. Thus, the surface density profile in the viscous region can be expressed as follows:
\begin{equation}
 \Sigma_\textrm{g,vis}=\Sigma_\textrm{g,vis} \tanh[(r-r_0)/(r_0h_0)].\\
\label{eq:density_visc}
\end{equation}

 For the outer region of the disk the corresponding profiles are given by
 
 \begin{equation}
 \begin{split}
 \Sigma_{\textrm{g},\textrm{irr}}= &2700\left(\frac{\dot{M_{\textrm{g}}}}{10^{-8} ~\textrm{M}_\odot~ \textrm{yr}^{-1}}\right)\left(\frac{M_{\star}}{\textrm{M}_{\sun}}\right)^{9/14}\left(\frac{\alpha_{\textrm{g}}}{10^{-3}}\right)^{-1}
     \\
     &\left(\frac{L_{\star}}{\textrm{L}_{\sun}}\right)^{-2/7}\left(\frac{r}{\textrm{au}}\right)^{-15/14}~\mathrm{g~cm^{-2}}
\end{split}
 \end{equation}

 \begin{equation}
 T_{\textrm{g},\textrm{irr}}= 150\left(\frac{M_{\star}}{\textrm{M}_{\sun}}\right)^{-1/7}\left(\frac{L_{\star}}{\textrm{L}_{\sun}}\right)^{2/7}\left(\frac{r}{\textrm{au}}\right)^{-3/7}~ \mathrm{K}
 \end{equation}

  \begin{equation}
 h_{\textrm{g},\textrm{irr}}= 0.024\left(\frac{M_{\star}}{\textrm{M}_{\sun}}\right)^{-4/7}\left(\frac{L_{\star}}{\textrm{L}_{\sun}}\right)^{1/7}\left(\frac{r}{\textrm{au}}\right)^{2/7}
 \end{equation}
 where $L_{\star}$ is the luminosity of the central object, and for simplicity,
 the disk was assumed to be vertical optically thin. 
 
 The boundary $r_\textrm{tran}$ that separates the viscous region from the irradiated region is given by
 
\begin{equation}
\begin{split}
    r_\textrm{tran}&=1.8\left(\frac{L_\star}{L_\odot}\right)^{-20/33} \left(\frac{M_\star}{M_\odot}\right)^{31/33} \left(\frac{\alpha}{10^{-3}}\right)^{-14/33} \\
    &\left(\frac{\dot{M_\textrm{g}}}{10^{-8}M_\odot yr^{-1}}\right)^{28/33}~\text{au}.
    \end{split}
\end{equation}
 
All the initial parameters were set at 1 Myr. We considered a central object with a mass $M_\star=0.08M_\odot$, which is close
to the substellar mass limit, and an evolving luminosity as predicted by 
the evolutionary models from \citet{Baraffe2015}.

The inner disk edge was set at $r_0=0.015 \textrm{au} \simeq 3R_{\star,0}$ with $R_{\star,0}$ the initial radius of the central object from the \citet{Baraffe2015} models and a corresponding aspect ratio $h_0=0.03$, calculated using $h_\textrm{g,vis}$ as $r_\textrm{tran}=0.086~au$ at 1 Myr. For the viscous coefficient, we used $\alpha_\textrm{g}=0.001$, which is a commonly  used value for disks around BDs \citep{Adame2011}.  
The gas accretion rate $\dot{M_\textrm{g}}$ was obtained from the fit proposed 
by \citet{Manara2012} which is based on a large sample of accreting
stars and BDs in the Orion nebular cluster. This approach was also adopted in 
\citet{Liu2020} to treat planet formation down to the substellar mass 
regime. Then, the 
evolution of $\dot{M_\textrm{g}}$ in time $t$ is given by

\begin{equation}
\begin{split}
    \log\left(\frac{\dot M_{\textrm{g}}}{\textrm{M}_\odot~ \textrm{yr}^{-1}}\right)=&-5.12-0.46\log\left(\frac{t}{\textrm{yr}}\right)-5.75\log\left(\frac{M_{\star}}{\textrm{M}_{\odot}}\right)\\
    &+1.17\log\left(\frac{t}{\textrm{yr}}\right)\log\left(\frac{M_{\star}}{\textrm{M}_{\odot}}\right).
\end{split}    
\end{equation}

Several observational results for young 
stellar populations show that for a given age, the number fraction 
of BDs harboring disks is higher than for low-mass stars (LMSs), which in turn is higher than the faction for more massive stars. For instance, at ages of 
$\sim7 \textrm{Myr}$, $\sim30\%$ of the BDs still retain their disks. For low-mass stars, this fraction is $5\%$ \citep[e.g.][]{Damjanov2007,Bayo2012,Riaz2012,
Dawson2013,Downes2015,ManzoM2020}. These results suggest that VLMSs 
and BDs could retain their primordial disks for longer times than 
more massive stars; the times may be up to several tens of million years. Then, we set a disk lifetime 
of $10~\textrm{Myr}$ for our disk model.
Although photoevaporation might cause some gas 
dispersion, the dominant dispersion process is the accretion onto the central object. As explained by 
\citet{stamatellos2015}, current uncertainties of how UV and X-ray emissions 
from BDs would affect their disks prevents an accurate estimate of the contribution 
of photoevaporation to the disk dispersion. Although is well known that this effect 
could occur in disks surrounding VLMSs \citep[e.g.][]{alexander2006}, their 
photoevaporation rates are estimated to be as low as $\sim10^{-11}~M_\odot/\textrm{yr}$ 
\citep{herczeg2007}. Even in the rough approximation of a constant photoevaporation 
rate of $10^{-11}~M_\odot/\textrm{yr}$ during the first $10~\textrm{Myr}$, the accretion onto the 
central object is always higher than this value in this period of time. We therefore ignored the effect of disk photoevaporation 
produced by radiation from the central object.

\section{Treatment of the disk-embryo interaction}
\label{sec:disk-embryo interaction}

In this section we describe and compare two different analytical prescriptions for the torques exerted by the gas-disk onto the embedded embryos. We compare them using test simulations.

\subsection{IDA20 and CN08 torque prescriptions.}
\label{sec:torques}

We analyzed the migration, and the eccentricity and inclination decay that embryos experience through their interaction with the gas component of the disk following the new
prescription from \citet[][from now on IDA20]{Ida2020} and the classic formulas from \citet[][from now on CN08]{CN08}. 
In both cases, the torques
that the gas exerts on the embryos were computed using the nonisothermal disk model from \citet{Paardekooper2010,Paardekooer2011}, which includes thermal and viscous diffusion. When a gravitational smoothing length of $b=0.4h_\textrm{g}$ is assumed, the total torque over each embryo is given by

\begin{equation}
    \Gamma_\textrm{total} = \Delta_\textrm{L}\Gamma_{\textrm{L}} +  \Delta_\textrm{C}\Gamma_{\textrm{C}},
\label{eq:torquetotal}
\end{equation}
where $\Delta_L$ and $\Delta_C$
are the reduction factors for noncircular or coplanar planetary orbits. The reduction factors differ in the CN08 and IDA20 prescriptions, as we
discuss in Sections \ref{sec:redfactIDA} and
\ref{sec:redfactCN08}. The factors $\Gamma_\textrm{L}$ and $\Gamma_\textrm{C}$ represent the Lindblad and corotation torques for a circular and coplanar motion, respectively, given by

\begin{equation}
    \Gamma_{\textrm{L}}=(-2.5-1.7\beta+0.1\delta)\frac{\Gamma_{0}}{\gamma_{\textrm{eff}}}
\end{equation}
and

\begin{equation}
\begin{split}
    \Gamma_{\textrm{C}}= & \Gamma_{\textrm{c},\textrm{hs},\textrm{baro}} F(p_{\nu})G(p_{\nu})+(1-K(p_{\nu})) \Gamma_{\textrm{c},\textrm{lin},\textrm{baro}} + \\ & \Gamma_{\textrm{c},\textrm{hs},\textrm{ent}}  F(p_{\nu}) F(p_{\chi})  \sqrt{G(p_{\nu})G(p_{\chi})} + \\ & \sqrt{(1-K(p_{\nu}))(1-K(p_{\chi}))} \Gamma_{\textrm{c},\textrm{lin},\textrm{ent}}.
    \end{split}
    \label{eq:corot_T}
\end{equation}
where $\Gamma_{\textrm{c},\textrm{hs},\textrm{baro}}$ and $\Gamma_{\textrm{c},\textrm{lin},\textrm{baro}}$ are barotropic terms related to the horseshoe drag and the linear corotation torque, respectively, and $\Gamma_{\textrm{c},\textrm{hs},\textrm{ent}}$ and $\Gamma_{\textrm{c},\textrm{lin},\textrm{ent}}$ are their corresponding nonbarotropic entropy counterparts. These terms are given by

\begin{equation}
\Gamma_{\textrm{c},\textrm{hs},\textrm{baro}}=1.1\left(1.5-\delta\right)\frac{\Gamma_{0}}{\gamma_{\textrm{eff}}},
\end{equation}

\begin{equation}
\Gamma_{\textrm{c},\textrm{lin},\textrm{baro}}=0.7\left(1.5-\delta\right)\frac{\Gamma_{0}}{\gamma_{\textrm{eff}}},
\end{equation}

\begin{equation}
\Gamma_{\textrm{c},\textrm{hs},\textrm{ent}}=7.9 \epsilon \frac{\Gamma_{0}}{\gamma_{\textrm{eff}}^{2}},
\end{equation}

\begin{equation}
\Gamma_{\textrm{c},\textrm{lin},\textrm{ent}}=\left(2.2-\frac{1.4}{\gamma_{\textrm{eff}}}\right) \epsilon \frac{\Gamma_{0}}{\gamma_{\textrm{eff}}},
\end{equation}
where the scaling torque is $\Gamma_0=\left(M_\textrm{p}/M_\star\right)^2\Sigma_\textrm{g}r^4h_{\textrm{g}}^{-2}\Omega_\textrm{k}^2$, with the angular Keplerian velocity $\Omega_\textrm{k}$. The negative of the entropy slope is $\epsilon=\beta-(\gamma-1)\delta$, with $\delta=-d\ln\Sigma_{\textrm{g}}/d\ln r$,  $\beta=-d\ln T_{\textrm{g}}/d\ln r$ and $\gamma=1.4$ the adiabatic index. In the viscous region, the parameter $\delta$ was calculated by considering the surface density profile for the viscous region from Eq. \ref{eq:density_visc}. The multiplied factor added to the original surface density viscous profile becomes relevant only for a region close to the inner edge of the disk up to $r \sim 0.016~au$. On the other hand, we note that in the irradiated region of the disk, $\delta$ and $\beta$ take values that make $\epsilon = 0$. In the irradiated region, the entropy contributions to the corotation torques are therefore null. The effective $\gamma_\textrm{eff}$ is given by

\begin{equation}
\gamma_{\textrm{eff}}=\frac{2Q \gamma}{\gamma Q+0.5\sqrt{2\sqrt{(\gamma^{2} Q^{2}+1)^{2}-16Q^{2}(\gamma-1)}+2 \gamma^{2} Q^{2} -2}}
\end{equation}
which is related to thermal diffusion by the coefficients $Q=2\chi/3h_\textrm{g}^3r^2\Omega_\textrm{k}$ and $\chi = 16 \gamma (\gamma-1) \sigma T_\textrm{g}^4/[3 \kappa (\rho_\textrm{g}h_\textrm{g} r \Omega_\textrm{k})^2]$, with $\sigma$ the Stefan-Boltzmann constant, $\kappa$ the gas opacity, and $\rho_\textrm{g}$ the volumetric gas density $\rho_\textrm{g}=\Sigma_\textrm{g}/
(H_\textrm{g} \sqrt{2\pi})$.

Additionally, the functions $F(p)$, $G(p)$, and $K(p)$ from Eq. \ref{eq:corot_T} are given by

\begin{equation}
F(p)=\frac{1}{1+\left(\frac{p}{1.3}\right)^2},
\end{equation}

\begin{equation}
G(p) = \left\lbrace
\begin{array}{lll}
\frac{16}{25}\left(\frac{45\pi}{8}\right)^{3/4} p^{3/2} & \textup{if } p<\sqrt{\frac{8}{45\pi}} \\
 1-\frac{9}{25}\left(\frac{8}{45\pi}\right)^{4/3} p^{-{8/3}} & \textup{if } p\geq\sqrt{\frac{8}{45\pi}} \end{array}
\right.,
\end{equation}

\begin{equation}
K(p) = \left\lbrace
\begin{array}{ll}
\frac{16}{25}\left(\frac{45\pi}{28}\right)^{3/4} p^{3/2} & \textup{if } p<\sqrt{\frac{28}{45\pi}} \\
 1-\frac{9}{25}\left(\frac{28}{45\pi}\right)^{4/3} p^{-{8/3}} & \textup{if } p\geq\sqrt{\frac{28}{45\pi}} \end{array}
\right..
\end{equation}
\\
The functions are evaluated in $p$, which takes the form of $p_\nu$, the saturation parameter associated with viscosity, or $p_\chi$, the saturation parameter related to thermal diffusion. Both parameters are given by 

\begin{equation}
p_{\nu}=\frac{2}{3}\sqrt{\frac{r^{2}\Omega_{\textrm{k}}x_{\textrm{s}}^{3}}{2\pi\nu}},
\end{equation}
where $x_\textrm{s}=(1.1/\gamma_\textrm{eff}^{0.25})\sqrt{M_\textrm{p}/(M_{\star}h_\textrm{g})}$ is the nondimensional half-width of the horseshoe region, and

\begin{equation}
p_\chi=\sqrt{\frac{r^2\Omega_\textrm{k}x_\textrm{s}^3}{2\pi\chi}}.
\end{equation}
Following \citet[][]{Coleman2014,Cossou2014,Izidoro2017,Carrera2018,Raymond2018}, we adopted a unique value for $\alpha$. Recent models \citep[e.g.,][]{Ida2018,Matsumura2021} include both a widen-driven disk accretion $\alpha$ for the inner disk profiles and a turbulent $\alpha$ to calculate the local planet-disk interactions. The same values we adopted for the accretion and turbulent $\alpha=10^{-3}$ also agree with one of the values proposed by \citet[][]{Matsumura2021}. An exploration of different values for the turbulent $\alpha$ is beyond the scope of this work.\\
All previous formulas that were used to calculate Lindblad and corotation torques were evaluated at the semimajor axis $a$ of the orbit of the embryo. The eccentricity $e$ and inclination $i$ values of the orbit of the embryo were included when the reduction factors were calculated.
 
\subsection{Reduction factors and acceleration from IDA20}\label{sec:redfactIDA}

 The new prescription from IDA20
 studies the gravitational interactions between the gas and the embryos on the basis of dynamical friction resulting in reduction factors given by
 \begin{equation}
    \Delta_\textrm{L}=\left(1+\frac{C_\textrm{P}}{C_\textrm{M}}\sqrt{e_\textrm{rat}^{2}+i_\textrm{rat}^{2}}\right)^{-1},
    \label{eq:deltaLida}
 \end{equation}
 
 \begin{equation}
     \Delta_\textrm{C}=\exp{\left(-\frac{\sqrt{e^{2}+i^{2}}}{e_\textrm{f}}\right)},
\label{eq:deltaCida}     
 \end{equation}
 where $C_\textrm{P}=2.5-0.1\delta+1.7\beta$, $C_\textrm{M}=6(2\delta-\beta+2)$, $e_\textrm{rat}=e/h_\textrm{g}$, $i_\textrm{rat}=i/h_\textrm{g}$ and $e_\textrm{f}=0.5h_\textrm{g}+0.01$.
 
In cylindrical coordinates $(r,\theta,z)$, the equations of motion for an embryo with a velocity $\vec{v}=(v_\textrm{r},v_\theta,v_\textrm{z})$ are given by

\begin{equation}
    \frac{d\vec{v}}{dt}=-\frac{v_\textrm{r}}{t_\textrm{e}}\hat{e_\textrm{r}}-\frac{(v_\theta-v_\textrm{k})}{t_\textrm{e}}\hat{e_\theta}-\frac{v_\textrm{k}}{2t_\textrm{a}}\hat{e_\theta}-\frac{v_\textrm{z}}{t_\textrm{i}}\hat{e_\textrm{z}},
\end{equation}
where $\hat{e_\textrm{r}}$, $\hat{e_{\theta}}$, and $\hat{e_\textrm{z}}$ are versors in the respective directions. The gas velocity is given by $\vec{v_\textrm{g}}=(0,(1-\eta)v_{\textrm{k}},0)$, with $v_\textrm{k}$ the Keplerian velocity, $\eta~\sim~ h_\textrm{g}^{2}$ and $\eta^{-1}t_\textrm{e}=t_\textrm{a}$. The variables $t_\textrm{a}$, $t_\textrm{e}$, and $t_\textrm{i}$ represent the damping timescales of the semimajor axis $a$, the eccentricity $e$, and the inclination $i$ of the orbit of the embryo, respectively. 
Considering that the embryo migration due to its interaction with the gas is nonisothermal and assuming the condition $i<h_\textrm{g}$, we can express the damping timescales as 

\begin{equation}
    t_{\textrm{a}} = -\frac{t_{\textrm{wave}}}{2h_{\textrm{g}}^{2}}\frac{\Gamma_0}{\Gamma_\textrm{total}},
\end{equation}

\begin{equation}
    t_\textrm{e} = \frac{t_\textrm{wave}}{0.78}\left[1+\frac{1}{15}(e_\textrm{rat}^{2}+i_\textrm{rat}^{2})^{3/2}\right],
\label{eq:teIDA}
\end{equation}

\begin{equation}
     t_\textrm{i} = \frac{t_\textrm{wave}}{0.544}\left[1+\frac{1}{21.5}(e_\textrm{rat}^{2}+i_\textrm{rat}^{2})^{3/2}\right],
\label{eq:tiIDA}
\end{equation}
where 
\begin{equation}
t_{\textrm{wave}} = \left(\frac{M_{\star}}{M_{\textrm{p}}}\right)\left(\frac{M_{\star}}{\Sigma_{\textrm{g}}r^{2}}\right)h_{\textrm{g}}^{4}\Omega_{\textrm{k}}^{-1},
\label{eq:twave}
\end{equation}
is the timescale from \citet{Papaloizou2000} and \citet{Tanaka2004}, in which all physical parameters are evaluated at the semimajor axis of the orbit. 

\subsection{Reduction factors and acceleration from CN08} \label{sec:redfactCN08}

The CNO8 prescriptions were obtained by fitting analytic formulas to hydrodynamic simulations of planets with eccentric and inclined orbits embedded in the gas disk. For the hydrodynamic simulations they treated the disk as a viscous fluid and preserved its mass by applying reflecting boundary conditions at the inner and outer boundaries. In this prescription the reduction factors are given by

\begin{equation}
    \Delta_\textrm{L}=\left[P_\textrm{e}+\frac{P_\textrm{e}}{\textbar {P_\textrm{e}} \textbar} \left\lbrace 0.07 i_\textrm{rat} + 0.085 i_\textrm{rat}^{4} - 0.08 e_\textrm{rat} i_\textrm{rat}^{2}\right \rbrace\right]^{-1},
\label{eq:deltaLCN08}
\end{equation}
where 
\begin{equation}
P_\textrm{e}=\frac{1+\left(\frac{e}{2.25h_\textrm{g}}\right)^{\textbf{6/5}}+\left(\frac{e}{2.84h_\textrm{g}}\right)^{6}}{1-\left(\frac{e}{2.02h_\textrm{g}}\right)^{4}} 
\end{equation}

\begin{equation}
 \Delta_\textrm{C}=\exp{\left(-\frac{e}{e_\textrm{f}}\right)}\left[1-\tanh{\left(i_\textrm{rat}\right)}\right].    
\label{eq:deltaCCN08} 
\end{equation}

From the comparison of the hydrodynamic simulations with N-body simulations, CN08 found that the acceleration for the embryos is given by

\begin{equation}
    \frac{d\vec{v}}{dt}=-\frac{\textbf{v}}{t_\textrm{m}}-2\frac{(\textbf{v}.\textbf{r})\textbf{r}}{r^{2}t_\textrm{e}}-2\frac{v_\textrm{z}}{t_\textrm{i}}\textbf{k},
\end{equation}
where $\textbf{r}$ and $\textbf{v}$ are the position and velocity vectors
of the embryo in Cartesian coordinates and $\textbf{k}$ is the versor in the z-direction.  Additionally, the migration, eccentricity and inclination damping timescales are given by

\begin{equation}
    t_\textrm{m}=-L\frac{{\Gamma_0}}{\Gamma_\textrm{total}},
\end{equation}

\begin{equation}
   t_\textrm{e}=\frac{t_\textrm{wave}}{0.78}\left(1-0.14e_\textrm{rat}^{2}+0.06e_\textrm{rat}^{3}+0.18e_\textrm{rat}i_\textrm{rat}^{2}\right),
\end{equation}

\begin{equation}
    t_\textrm{i}=\frac{t_\textrm{wave}}{0.544}\left(1-0.3i_\textrm{rat}^{2}+0.24i_\textrm{rat}^{3}+0.14e_\textrm{rat}^{2}i_\textrm{rat}\right),
\end{equation}
where $L=M_\textrm{p}\sqrt{G M_\star a (1-e^{2})}$ is the orbital angular momentum of the embryo, $M_\textrm{p}$ is its mass, $G$ is the gravitational constant, $t_\textrm{wave}$ is given by Eq. \ref{eq:twave} and the total torque $\Gamma_\textrm{total}$ includes the reduction factors given by Eq. \ref{eq:deltaLCN08} and Eq. \ref{eq:deltaCCN08}.

It is important to point out that in CN08, the associated accelerations are related with $t_\textrm{m}$, $t_\textrm{e}$, and $t_\textrm{i}$, while in IDA20, they are associated with $t_\textrm{a}$, $t_\textrm{e}$, and $t_\textrm{i}$. We clarify that the timescales are related by

\begin{equation}
    t_\textrm{m}=\left[\frac{1}{2}t_\textrm{a}^{-1}-
    \frac{e^{2}}{1-e^{2}}t_\textrm{e}^{-1}-\textrm{i}^{2}t_\textrm{i}^{-1}\right]^{-1}.
    \label{eq:tm}
\end{equation}

The migration timescale $t_\textrm{m}$ does not represent the actual evolution of the semimajor axis. Thus a change in the orbital angular momentum $L$ could happen both with a change in semimajor axis or with a change in the orbital eccentricity of a planet. This indicates that an inward migration is not always associated with a decay in semimajor axis if the orbit is noncircular. 

\subsection{Comparison of type I migrations from IDA20 and CN08}
\label{sec:mig-comparison}

The migration and orbital decay experienced by the embryos due to the gas were modeled by IDA20
and CN08. Both teams considered the same estimates for the Lindblad and corotation torques, but differed in the reduction factors when considering a noncircular or coplanar orbit, the damping timescales for $a$, $e$, and $i$ and the accelerations terms.
We evaluated the consistency of the two prescriptions by comparing the
corresponding total torques for an embryo in two different scenarios: when its orbit is circular and coplanar with the disk midplane, and when it is not. In both
cases we used the disk model described in Section \ref{sec:diskmodel}. 

\begin{figure*}
    \centering
    \includegraphics[width=0.54\textwidth]{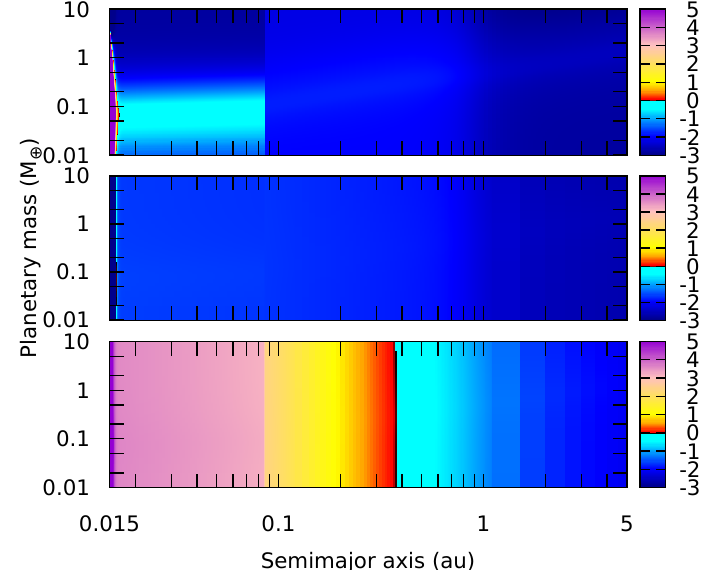}
    \includegraphics[width=0.355\textwidth]{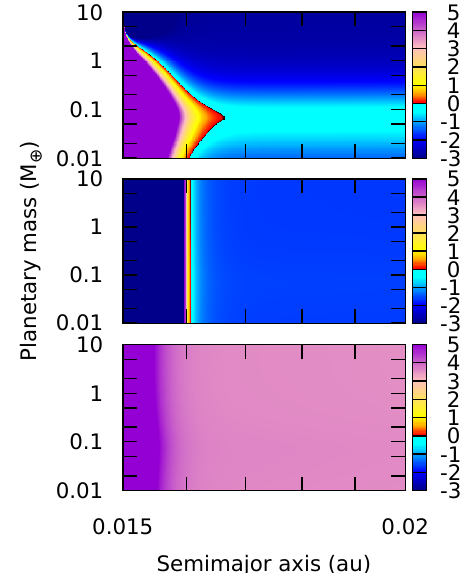}
     \caption{Maps of the total normalized torque $\Gamma_\textrm{total}/\Gamma_0$ 
    exerted by the gas onto embryos in a $1~\textrm{Myr}$ old disk for: coplanar and circular orbits (top panels), orbits with $e=0.1$ following the prescriptions from IDA2020 (middle panels), and orbits with $e=0.1$ following the prescriptions from CN08 (bottom panels). The right panels show zoom-ins of the inner part of the disk.}
    \label{fig:mapatorques1M}
\end{figure*}

Fig. \ref{fig:mapatorques1M} shows the maps of $\Gamma_\textrm{total}$ 
normalized by $\Gamma_0$ for embryos with masses $M_{\textrm{emb}}$ and semimajor 
axis $a$ within the ranges $M_\leftmoon<M_{\textrm{emb}<10~M_\oplus}$ and 
$0.015<a/\textrm{au}<5$, and setting the disk parameters for an age of $1~\textrm{Myr}$.
The left panel shows the complete range of $a$, and the right panel focuses on 
the inner part $0.015<a/\textrm{au}<0.02$.

The top panels show the circular and coplanar case, in which the reduction factors 
$\Delta_L$ and $\Delta_C$ equal unity and both prescriptions result in the 
same $\Gamma_\textrm{total}$. In this case, the total torque becomes positive for embryos closer to the inner edge and for masses 
$M_{\textrm{emb}}<3~M_\oplus$, while $\Gamma_\textrm{total}\sim0$ only 
for embryos at some particular distances close to the inner edge of the disk, which depends on the mass of the embryo. For the remaining combinations of $a$ and $M_\textrm{emb}$ the 
total torque $\Gamma_\textrm{total}$ remains negative. However, it became even more negative for $r>r_\textrm{tran}$ and $M_\textrm{emb}>0.3~M_\oplus$. 
The middle panels shows the results considering coplanar orbits with $e=0.1$ 
and $\Gamma_\textrm{total}$ calculated using $\Delta_L$ and $\Delta_C$ from 
the IDA20 prescription. In this case $\Gamma_\textrm{total}$ is always 
negative (even more negative for $r>r_\textrm{tran}$), except for a narrow region at $a\sim0.016~\textrm{au}$ where it becomes positive even close to zero. This small region of positive torque values is produced by the maximum of the density profile of the gas and the inner negative values for the torques are due to the trap considered close to the inner edge. As we described in Sec. \ref{sec:redfactIDA}, $\Delta_L$ under the IDA20 prescription includes the parameters $\delta$ and $\beta$, which are related to the disk density profiles and lead to differences when either the eccentricity or the inclination of the planet is non-negligible. The bottom panel shows the $\Gamma_\textrm{total}$ calculated for a coplanar orbit with 
$e=0.1$ using $\Delta_L$ and $\Delta_C$ from CN08. In this case, $\Gamma_\textrm{total}$
changes from positive values in the inner region of the disk to negative values 
in the outer disk. The values are almost zero around $a\sim0.4~\textrm{au}$. The region closest to the inner edge has the highest positive torque values, and it decreases with r even more after $r>r_\textrm{tran}$. The increase in $e$ produces positive values of $\Gamma_\textrm{total}$ for a wide range of radial distances under the CN08 treatment. However, there is no particular treatment for the trap close to the inner edge because the factors $\Delta_L$ and $\Delta_C$ are not related with 
the disk profiles, just with the value of the eccentricity and inclination.

We found that the addition of an orbital inclination $i \leq h_\textrm{g}$ results in similar torque patterns, but they are closer to zero than those in the previous analysis. As $i$ 
increases, $\Gamma_\textrm{total}$ becomes even closer to zero, as expected because the embryo orbits are further from the midplane of the disk most of the time.

Finally, we explored how the $\Gamma_\textrm{total}$ maps change throughout the gas-disk lifetime. We found that at $t\sim3~\textrm{Myr}$ and later, the 
$\Gamma_\textrm{total}$ remained negative for the complete ranges of $M_{emb}$
and $a$ for both circular and coplanar orbits and also for orbits with $e=0.1$ if it follows the IDA20 prescription. On the other hand, for an orbit with $e=0.1$ under the CN08 treatment, an analogous pattern in the $\Gamma_\textrm{total}$ map remained, but the inner positive torques decreased in absolute value and the outer positive torques were extended up to higher values of $a$ until they reached $a\sim0.8$ $\textrm{ua}$ at $t~=10~\textrm{Myr}$.

To summarize, the $\Gamma_\textrm{total}$ maps from the two prescriptions differ significantly in 
the case of noncircular orbits. We therefore searched for the differences in the evolution 
history paths of the embryos and the final architectures of the planetary 
systems when we applied one or the other prescription.

\subsection{Test of numerical simulations}

We modified the well-known $\textsc{Mercury}$ code \citep[][]{Chambers1999} by 
adding the torque prescriptions from IDA20 and CN08 and the disk model discussed in Section  \ref{sec:diskmodel} to our previous modification including tidal 
and general relativistic effects \citep[][]{Sanchez2020}. Then, we tested the agreement within the
external forces and the damping timescales by analyzing the orbital evolution from a set of $N$-body simulations of a planet with different initial parameters that followed either IDA20 or CN08.

\begin{figure}
    \centering
    \includegraphics[width=8cm,height=9cm]{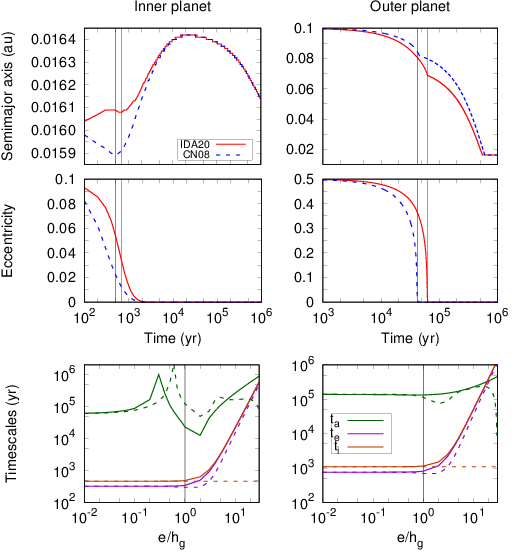}
    \caption{First $1~\textrm{Myr}$ evolution of $a$ and $e$, 
    and the absolute values of 
    $t_a$, $t_e$ and $t_i$ for an inner planet (left panels) initially located at
    $a=0.016~\textrm{au}$ with $e=0.1$ and an outer planet (right panels) initially 
    located at $a=0.1~\textrm{au}$ and $e=0.5$. The planet masses are 
    $M_\textrm{P}=M_\textrm{Mars}$. Solid lines indicate the results following the 
    prescriptions from IDA20, and dotted lines show the corresponding results from CN08. 
    The vertical lines indicate the moment at which $e=h_\textrm{g}$, which separates 
    the subsonic ($e<h_\textrm{g}$) and the 
    supersonic ($e>h_\textrm{g}$) regimes.}
    \label{fig:aet_evolution}
\end{figure}

Fig. \ref{fig:aet_evolution} shows the first $1~\textrm{Myr}$ evolution
of $a$, $e$, and the absolute values of $t_a$, $t_e$ and $t_i$ for an inner planet 
initially located in an orbit with $a=0.016~\textrm{au}$ and $e=0.1$ and an outer 
planet initially located in an orbit with $a=0.1~\textrm{au}$ and $e=0.5$. In both 
cases the planets have masses $M_P=M_\textrm{Mars}$ and the simulations were performed 
twice following the prescriptions from IDA20 and CN08. We indicate the separation between the subsonic ($e < h_\textrm{g}$) and supersonic ($e > h_\textrm{g}$) regimes.

Regarding the inner planet evolution, if we follow IDA20, while $e < h_\textrm{g}$, the embryo first moves away from the star and then moves slowly inward until $e=h_\textrm{g}$, from where we can separate the direction of migration into two: first the planet moves outward until $\Gamma_\textrm{total}=0$ when the planet starts moving inward because of the change in the torque sign. On the other hand, when we follow CN08, while $e<h_\textrm{g}$, the planet moves inward until it reaches $e=h_\textrm{g}$, at which time it reproduces the same migration direction as for IDA20. When $e$ is non-negligible, $\Gamma_\textrm{total}$ does not give the direction of evolution of the semimayor axis when the planet follows CN08, but it does when the planet follows IDA20 (see the torques in Fig. \ref{fig:mapatorques1M}). This means that the CN08 prescription involves $t_\textrm{m}$ instead of $t_\textrm{a}$ in the  acceleration expressions of the planets, which can differ in sign as $t_\textrm{m}$ is related to both $t_\textrm{a}$ and $t_\textrm{e}$ (see Eq. \ref{eq:tm}). On the other hand, when the embryo orbit starts to circularize, an inward migration is represented by $\Gamma_\textrm{total}<0$ and an outward migration by $\Gamma_\textrm{total}>0$, in both prescriptions, thus $t_\textrm{m}$ and $t_\textrm{a}$ preserve their sign.

\begin{figure}
    \centering
    \includegraphics[width=7.5cm,height=5cm]{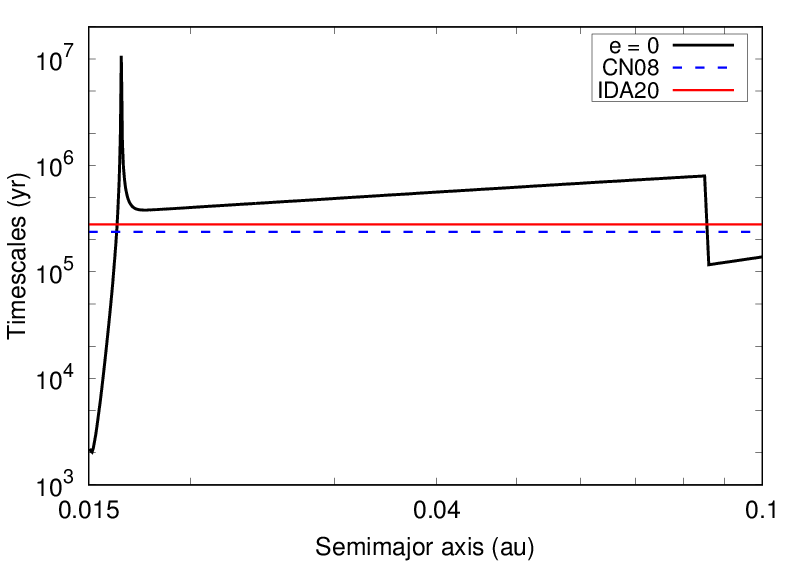}
    \caption{Timescale $t_\textrm{a}$ as a function of $a$ for a planet with a mass $M_\textrm{P}=\textrm{Mars}$. 
    The black line indicates the case of a circular and coplanar orbit. The red and blue lines indicate a planet initially located at $a=0.1~\textrm{au}$ with $e=0.5$ following the IDA20 and CN08 prescriptions, respectively. All cases consider disk parameters set to $1~\textrm{Myr}$.}
    \label{fig:ta_comparison}
\end{figure}

By following either of the two prescriptions, the outer planets moves inward. 
As in the case of the inner planet, the sign of $\Gamma_\textrm{total}$
does not agree with the direction in which $a$ evolves for a noncircular orbit when we use the CN08 prescription. On the other hand, the planet migration under the IDA20 
treatment coincides at any position with a $\Gamma_\textrm{total}<0$, in agreement with an inward migration. For the outer planets, the semimajor axis decreased considerably up to one order of magnitude during the integration time. The sinks in the evolution of $a$ from both prescriptions when the planet reached 
$e = h_\textrm{g}$ came from the variation in $t_\textrm{a}$ that the planet experienced while migrating inward. Fig. 
\ref{fig:ta_comparison} shows the initial values of $t_\textrm{a}$ from CN08 and 
IDA20, which were lower than the values they took when the orbit became 
circular. In the case of a circular orbit, $t_\textrm{a}$ presents two peaks, one 
related to a distance that is close to the inner edge of the disk ($a \sim 0.0165~\textrm{au}$) 
where $\Gamma_\textrm{total}$ changes its sign, and the other related to 
the distance $a=r_\textrm{trans}\sim 0.8~\textrm{au}$ that separates the viscous and 
irradiated zones of the disk where $\Gamma_\textrm{total}$ starts to 
decrease as shown in Fig. \ref{fig:mapatorques1M}.

For the inner and outer planet simulations, $t_a$ shows the larger difference within the two prescriptions in the supersonic regime. Oscillations in their absolute values are visible that are associated with the calculation of $\Gamma_\textrm{total}$, which differs from one prescription to 
the other. On the other hand, when the orbit is quasi-circular, both timescales are similar. 

When the orbit is quasi-circular, the timescales $t_e$ and $t_i$ are equivalent in the two prescriptions. When the orbit is eccentric while $t_i$ remains constant for CN08, it follows the same increment as $t_e$ for IDA20, because one depends on the other, as shown in Eq. \ref{eq:teIDA} and Eq. \ref{eq:tiIDA}.

Finally, we explored the effect of orbital inclination through a set of new simulations 
for an orbit initially inclined by an angle $i \leq h_\textrm{g}$. We found no modification 
in the evolution for $a$, $e$, $i$, $t_i$, and $t_a$ while the timescale $t_e$ differed 
slightly within the two prescriptions, although these differences did not produce any 
considerable change in the orbital evolution or migration directions.\\

Our analysis shows important differences in the orbital evolution of a planet for CN08 or IDA20. We therefore decided to develop two sets of simulations for a detailed study of the impact of the two prescriptions on the formation and evolution of a planetary system around an object close to the substellar mass limit.

\section{Simulations of planetary system formation}
\label{sec:simulations}

We present the scenario of planetary system formation starting with a sample of embryos orbiting an evolving central object with a mass of $0.08~M_\odot$. We developed a set of 23 $N$-body simulations following the prescription of IDA20 and 
another 23 simulations following CN08 with our modified version of the $\textsc{Mercury}$ 
code. As external forces, we incorporated the acceleration corrections generated by 
the interaction within the gas-disk and the embryos of the two prescriptions, as well as those produced by tidal and general relativistic effects which are relevant during 
rocky planet formation at the substellar mass limit, as we showed in \cite{Sanchez2020}. 
We also included the central object contraction and rotational period evolution as well as 
a fixed pseudo-synchronization period for embryos.

\subsection{Tidal and general relativistic effects}

We incorporated tidal effects following the equilibrium tide model from \citet{Hut1981} 
and \citet{Eggleton1998}. We included tidal distortions and dissipation terms, considering 
the tide raised by the central object on each embryo and by each embryo on the 
central object and neglected the tide between embryos as follows,

\begin{equation}
  \textbf{f}_\omega = -3\frac{\mu}{r^8}\left[k_{2,\star}\left(\frac{M_\mathrm{p}}{M_\star}\right)R_\star^5 + k_{2,\mathrm{p}}\left(\frac{M_\star}{M_\mathrm{p}}\right)R_\mathrm{p}^5\right] \textbf{r},
\end{equation}

\begin{align*}
\textbf{f}_{\textrm{ae}} = & -3\frac{\mu}{r^{10}} \left[
  \frac{M_{\textrm{p}}}{M_\star} k_{2,\star} \Delta \mathrm{t}_\star R_\star^{5}\left(2\textbf{r}(\textbf{r} \cdot \textbf{v}) + r^{2}(\textbf{r} \times \Omega_\star + \textbf{v})\right)\right]
\end{align*}
\begin{equation}
-3\frac{\mu}{r^{10}} \left[\frac{M_\star}{M_{\textrm{p}}}k_{2,\textrm{p}} \Delta \mathrm{t}_{\textrm{p}} R_\textrm{p}^{5} 
       \left(2\textbf{r}(\textbf{r}\cdot\textbf{v}) + r^{2}(\textbf{r} \times \Omega_\textrm{p} + \textbf{v})\right)\right],
\end{equation}
where $k_{2,\star}=0.307$ and $k_{2,\textrm{p}}=0.305$ are the potential Love numbers of 
degree 2 of the star and the embryos, respectively. For the star we assumed the Love number of an object with a mass at the substellar mass limit, and for the embryos, we assumed the Love number estimated for the Earth \citep{Bolmont2015}. The variable \textbf{r} is the position vector of the embryo with respect to the central object,
$\mu=G(M_\star + M_\textrm{p})$, $G$ is the gravitational constant, and $M_\star$, $R_\star$, 
$M_\textrm{p}$ and $R_\textrm{p}$ are the masses and radius of the star and the
protoplanetary embryo, respectively, under the approximation that these objects can instantaneously 
adjust their equilibrium shapes to the tidal force and considering only up to second-order harmonic distortions \citep{Darwin1908}. The variable $\textbf{v}$ is the velocity vector of the embryo with respect to the central star, $\Delta \mathrm{t}_\star$ 
and $\Delta \mathrm{t}_\mathrm{p}$ are the time-lag model constants 
for the star and the protoplanetary embryo, respectively. The factors 
$k_{2,\star}\Delta \mathrm{t}_\star$ and $k_{2,\mathrm{p}}\Delta \mathrm{t}_\mathrm{p}$, 
are related to the dissipation factors by
\begin{equation}
  k_{2,\mathrm{p}} \Delta \mathrm{t} _\mathrm{p} =  \frac{3R_\mathrm{p}^5\sigma_\mathrm{p}}{2G} \\
   k_{2,\star} \Delta \mathrm{t}_{\star} = \frac{3R_{\star}^5\sigma_{\star}}{2G}
\end{equation}  
with the dissipation factor for each protoplanetary embryo 
$\sigma_\mathrm{p}=8.577\times10^{-43}~\mathrm{k^{-1} m^{-2} s^{-1}}$, which is the same dissipation factor as estimated for the Earth \citep{Neron1997}, 
and the dissipation factor of the central object is 
$\sigma_\star=2.006\times10^{-53}~\mathrm{k^{-1} m^{-2} s^{-1}}$ 
\citep{Hansen2010}.\\ 
We included the rotational evolution ($\Omega_\star$) 
and contraction of the central object ($R_\star$) from \citet{Bolmont2011} and \citet{Baraffe2015}, and fixed 
each embryo at pseudo-synchronization ($\Omega_\textrm{p}$) following \cite{Hut1981}. Thus, in a heliocentric reference frame, tidal interactions lead the precession of the argument of periastron $\omega$,
as well as the $a$ and $e$ decays.\\
We also incorporated the acceleration corrections 
associated with the precession of periastron caused by the central object as derived from the General 
Relativity Theory (GRT) \citep{Einstein1916} as follows,
\begin{equation}
  \textbf{f}_{\mathrm{GR}} = \frac{GM_\star}{r^3c^2}\left[\left(\frac{4GM_\star}{r} - \textbf{v}^2\right)\textbf{r}+4(\textbf{v}.\textbf{r})\textbf{v}\right],
  \label{eq:grav}
\end{equation}
with $c$ the speed of light. Eq. \eqref{eq:grav} was proposed by 
\citet{Anderson1975}, who used the parameterized post-Newtonian theories and reported a relative correction associated with two parameters $\beta$ and 
$\gamma$, which are equal to unity in the GRT case. We refer to 
\citet{Sanchez2020} for a detailed description of tidal and relativistic 
corrections and their associated orbital decay timescales.

\subsection{Initial distribution of embryos}

The initial spatial distribution of embryos extends from $r_\textrm{ice}<r<r_\textrm{final}$, 
where $r_\textrm{ice}=0.23~\textrm{au}$ is the location of the snow line at $1~\textrm{Myr}$ and 
$r_\textrm{final}=5~\textrm{au}$ as in \citet[][]{Coleman2019}. The location of the snow line 
was computed following the parameterization from \citet{Ida2016} given by 
$r_\textrm{ice} \sim max(r_\textrm{ice,vis},r_\textrm{ice,irr})$, with
\begin{equation}
r_\textrm{ice,vis}=1.2\left(\frac{M_\star}{M_\odot}\right)^{1/3} \left(\frac{\dot{M_{\textrm{g}}}}{10^{-8}M_\odot yr^{-1}}\right)^{4/9} \left(\frac{\alpha_{\textrm{g}}}{10^{-3}}\right)^{-2/9}~au,
\end{equation}
and
\begin{equation}
r_\textrm{ice,irr}=0.75\left(\frac{M_\star}{M_\odot}\right)^{-1/3} \left(\frac{L_\star}{L_\odot}\right)^{2/3}~\textrm{au},   
\end{equation}
where $\dot{M_\textrm{g}}$ and $L_\star$ are evaluated at $1~\textrm{Myr}$. The first embryo was located at $a=r_\textrm{ice}$ while the location of the remaining consecutively embryos was calculated with $a_\textrm{i+1}=a_{i}+\Delta R_{\textrm{Hill}}$, assuming $\Delta$ to be a randomly integer number between 5 and 10 and
$R_\textrm{Hill}=a(2M_\textrm{emb}/3 M_\star)^{1/3}$, with $i=1,2,etc$.\\
Our simulations are intended to explore the rocky 
planet formation from an embryo population and do not include  
earlier formation stages such as pebble accretion or planetesimals. We therefore set an initial sample of already formed embryos with masses $M_\textrm{emb}= 0.16M_\oplus$ ($\sim 1.5 M_\textrm{Mars}$) comparable to those used in previous work on planet formation at the substellar mass 
limit \citep[e.g.,][]{Coleman2019}.

The number of embryos was computed from the total mass of solids 
$M_\textrm{solid}$ as

\begin{equation} 
 \label{eq:mdust}
 M_\textrm{solid} = 2\pi  \int_{r_\textrm{ice}}^{r_\textrm{final}} r \Sigma_\textrm{solid} dr,
 \end{equation}

where

\begin{equation}
    \Sigma_\textrm{solid}=\left\lbrace
    \begin{array}{cc}\Sigma_\textrm{g,vis} z_0 \eta_\textrm{ice} &~if~ r < r_\textrm{tran}, \\
    \\
    \Sigma_\textrm{g,irr} z_0 \eta_\textrm{ice} &~if~ r > r_\textrm{tran},\end{array}\right.
\end{equation}
where $\Sigma_\textrm{solid}$ is the solid surface density profile,
$z_0=0.0153$ is the primordial solar abundance of heavy elements 
\citep[][]{Lodders2009}, and $\eta_\textrm{ice}$ represents the increase in the 
amount of solids due to the condensation of water at $r>r_\textrm{ice}$ whose 
values from \citet{Lodders2003} are $\eta_\textrm{ice}=1~if~ r<r_\textrm{ice}$ and 
$\eta_\textrm{ice}=2~ if~ r>r_\textrm{ice}$. The factor $2$ in $\Sigma_\textrm{solid}$ 
is related to the water radial distribution, so that all the 
bodies located outside the snow line have $50\%$ of water in mass. Then, by solving Eq. \ref{eq:mdust} we obtained $M_\textrm{solid}\sim 7 M_\oplus$. We considered that 
$\sim 30 \%$ of the mass is the dust that causes the opacity in the disk, which is 
representative of the typical values of dust masses that have been found in disks around 
low-mass objects at early ages \citep[e.g.,][]{Duong2018}. Then the initial mass in embryos is $\sim 4.5 M_\oplus$ and the total number of initial embryos is $28$. We 
neglected the mass of solids between $r_\textrm{init}$ and $r_\textrm{ice}$ because is 
negligible in comparison with the mass obtained from solving Eq. \ref{eq:mdust}.

We considered that all embryos initially have $e<0.02$ and $i<0.5^\circ$, and the initial
values of the argument of periastron $\omega$, longitude of 
the ascending node $\Omega$, and mean anomaly $M$, were randomly determined for each 
embryo from uniform distributions between $0^\circ$ and $360^\circ$.

\subsection{Characterization of $N-$body simulations}

To develop our simulations, we chose the hybrid integrator, which uses a second-order 
symplectic algorithm to treat interactions between objects with separations greater than 
$3R_\textrm{Hill}$ and the Bulirsch-St\"oer method to solve close encounters. The time step we adopted corresponds to $1/30$~th of the orbital period of the innermost body of the 
simulations, which is $0.08$~days. 
We considered an embryo as ejected from the system when it reached a distance $r>100~\textrm{au}$,
and we considered that the embryo had collided with the central star when it was closer than 
$0.0045~\textrm{au}$, which corresponds to the maximum radius of the star. We fixed this value for the entire simulation in order to avoid any numerical error for small-perihelion 
orbits. All simulations ran for $100~\textrm{Myr}$ in order to analyze the dynamic of 
planetary systems well after the gas has dissipated from the disk at $10~\textrm{Myr}$.

\section{Results}
\label{sec:results}

In this section we analyze the gas effects on the dynamical evolution of 
embryos during the gas-disk lifetime regarding the prescriptions from IDA20 
and CN08 independently, and also the dynamics of planetary systems after the gas 
has been dissipated from the disk. We show the final architectures of planetary 
systems, focusing on the close-in planet population, and compare it with 
observational results. 

\subsection{Gas stage}\label{sec:gasstage}

\begin{figure*}
    \centering
    \includegraphics[width=18cm,height=11cm]{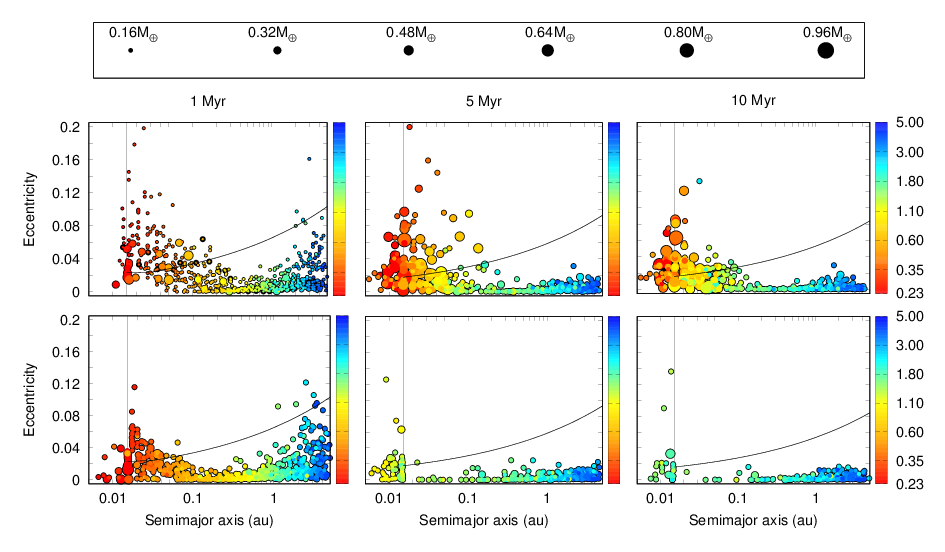}
    \caption{Semimajor axis as a function of eccentricity for embryos at different times 
    before the complete dissipation of the gas at $10~\textrm{Myr}$. The color scale represents the initial 
    $a$ of each embryo, and the sizes indicate their masses. The top and bottom panels
    show the results from simulations including the prescriptions from IDA20 and CN08, respectively.
    The black lines indicates $e = h_\textrm{g}$, and the vertical lines the inner edge of the 
    disk $(r_\textrm{inner}=0.015~\textrm{au})$.}
    \label{fig:ae_gas}
\end{figure*}
\begin{figure}
    \centering
    \includegraphics[width=8.5cm,height=7cm]{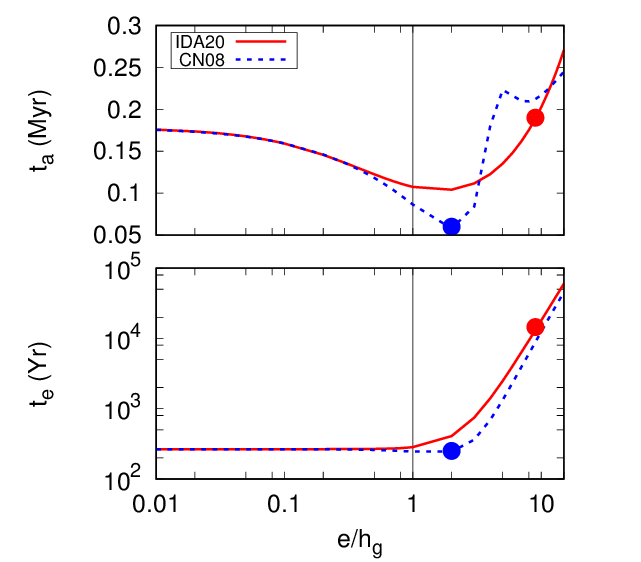}
    \caption{Semimajor axis migration timescales $t_a$ (top panel) and eccentricity damping timescales $t_e$ (bottom panel) as a function of $e/h_\textrm{g}$. The red and blue lines indicates the prescriptions from IDA20 and CN08, respectively. Both cases consider an inner planet with mass  $M_{Mars}$ located at $a=0.02~\textrm{au}$. The red dot indicates $e=0.2$ from the IDA20 prescription, and the blue dot shows $e=0.05$ from CN08. The solid black line represents the limit between subsonic and supersonic regime.}
    \label{fig:comp_escalas}
\end{figure}

As discussed in Section \ref{sec:disk-embryo interaction}, the gas disk exerts 
torques on the embryos, allowing them to migrate. The direction of the 
migration depends on each prescription for the disk model used, as well as on 
the physical and orbital parameters of the embryos. The timescales for 
orbital decay in both treatments are comparable when the planetary orbit is 
quasi-circular and coplanar with the midplane of the disk, but they differ from each other 
when the orbit is eccentric. This 
discrepancy has an important effect on the dynamic of embryos, as 
we discuss below.

Fig. \ref{fig:ae_gas} shows $e$ as a function of $a$ for the embryos 
that have survived in the disk at $1~\textrm{Myr}$, $5~\textrm{Myr}$, and $10~\textrm{Myr}$ during the gas-disk lifetime, distinguished by their initial semimajor axis and mass. In the close-in population ($a<0.1~\textrm{au}$), the embryos decreased their $e$ and migrated inward faster under the CN08 prescription 
than under the IDA20 treatment. As an example of this behavior, Fig. \ref{fig:comp_escalas} shows the $t_a$
and $t_e$ damping timescales for an embryo with a mass of $M_{Mars}$
located at $a=0.02~\textrm{au}$ for the IDA20 and CN08 
prescriptions.\\
The eccentricity damping timescales are shorter in the CN08 regime. Thus the planets involved have lower eccentricities than those ones under the IDA20 treatment (see Fig. \ref{fig:ae_gas}). This difference in eccentricity in the supersonic regime causes that for a given time, at a particular location of the planet, the semimajor axis damping timescale is higher for an IDA20 embryo than for a CN08 embryo, as shown in Fig. \ref{fig:comp_escalas}.
This explains why the inner embryo population migrated faster and left the region $0.015 < a/\textrm{au} < 0.1$. Only one embryo survived in $\sim 17 \%$ of the CN08 simulations in quasi-circular and coplanar orbits, but more than one embryo survived in this region in all IDA20 simulations, in some cases, in more eccentric orbits (see Fig. \ref{fig:ae_gas}).\\

\begin{figure}
    \centering
    \includegraphics[width=8cm,height=6cm]{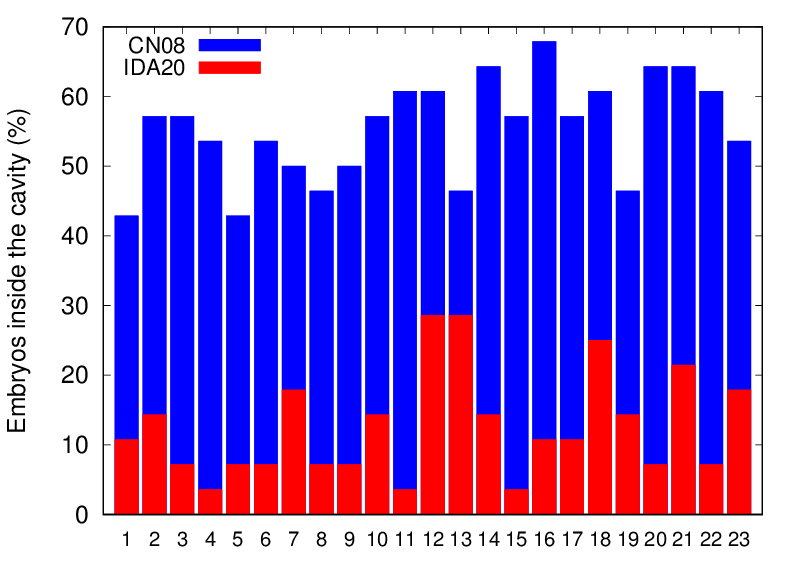}
    \caption{Percentage of the initial number
   of embryos that entered the cavity ($a<0.015~au$) during the gas stage. Each bin indicates a single simulation. Red bars represent simulations 
   using IDA20, and blue bars show simulations using CN08.}
    \label{fig:inside_cavity}
\end{figure}

Fig. \ref{fig:inside_cavity} shows the fraction over the initial number of embryos that ended up inside the cavity ($a < 0.015~\textrm{au}$) during the gas stage for both prescriptions. In all simulations, between $\sim$ 45$\%$ and 70$\%$ of the total number of initial embryos entered the cavity under the CN08 treatment, while just between $\sim$ 5$\%$ and 30$\%$ of them did so under the IDA20 prescription. A greater number of embryos entered the cavity under the CN08 treatment for the fast migration and the strong total torque at the inner edge of the disk that the embryos experienced, especially during the first 5 Myr of the disk lifetime. \\

Approximately 10 $\%$ of the embryos that entered the cavity under the CN08 prescription collided, and then the remaining 90 $\%$ collided with the central object. Under the IDA20 treatment, $\sim$ 10 $\%$ of the embryos that entered the cavity collided in only $\sim$ 30 $\%$ of the simulations, while in the remaining $\sim$ 70 $\%$ of the simulations, all the embryos that entered the cavity collided directly with the central object during the integration time. \\

\begin{figure}
    \centering
    \includegraphics[width=8cm,height=6cm]{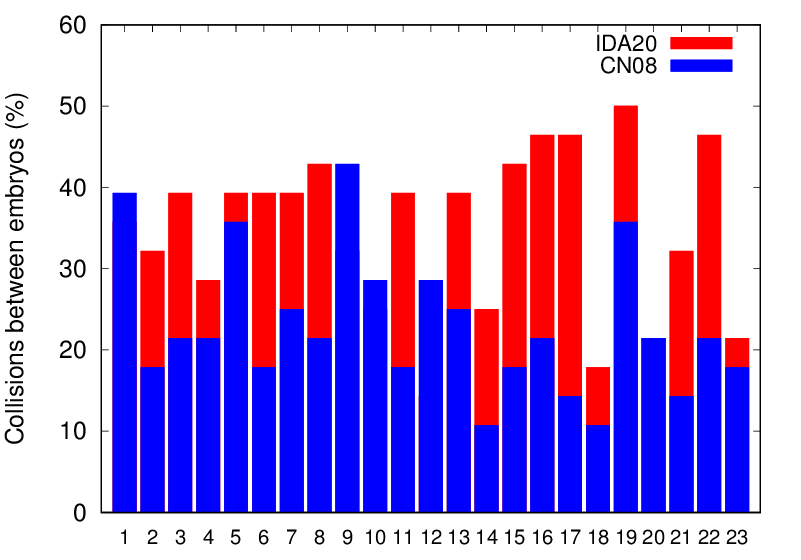}
    \caption{Percentage of the initial number
   of embryos that collided during the gas stage. Colors are as in Fig. \ref{fig:inside_cavity}}.
   \label{fig:col_emb_gas}
\end{figure}

Fig. \ref{fig:col_emb_gas} shows the percentage of the initial number of embryos that collided during the disk lifetime ($10~\textrm{Myr}$) for every simulation of each of the two prescriptions. The collisions were more frequent for embryos in simulations including the IDA20 prescription $(\sim 78\%)$ than those in CN08 $(\sim 22\%)$. In both cases, about $98\%$ of the collisions occurred among embryos with $a < 0.1~\textrm{au}$. Under the CN08 prescription, all the embryos that collided inside the disk entered the cavity. On the other hand, under the IDA20 prescription, this occurred for all of the embryos that collided inside the disk, in later stages during the gas-disk lifetime, given the slow migration rate under this treatment, and thus they survived in the planetary systems.

\begin{figure}
    \centering
    \includegraphics[width=9cm,height=7cm]{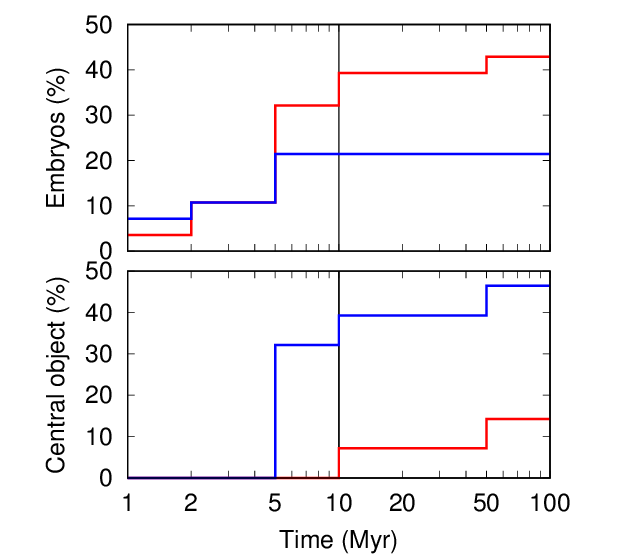}
    \caption{Medians of accumulated collisions of embryos (top panel) and of embryos that collided with the central object (bottom panel) of all IDA20 (red lines) and CN08 simulations (blue lines) before and after the dissipation of the gas-disk at 10 $\textrm{Myr}$ (vertical line).}
    \label{fig:medians}
\end{figure}

For a detailed analysis of the collision history of embryos during the gas-disk lifetime, Fig. \ref{fig:medians} shows the medians of the accumulated collisions of embryos and with the central object, during and after the gas dissipated from the disk, considering all the simulations for each prescription. During the first $3~\textrm{Myr}$, no collisions with the central object occurred in the IDA20 
simulations, while less than $10 \%$ of these collisions occurred in $\sim 17 \%$ of the CN08 simulations, even though the median value is equal to zero. Furthermore, a median of $\sim 10 \%$ of the embryos collided in the simulations of each treatment. Later, between 3 and 5~Myr, the IDA20 simulations have a median of collisions among embryos of $\sim 32 \%$ and still a median of $\sim 0 \%$ of collisions with the central object, while the CN08 simulations present a median of $\sim 21 \%$ and 
$\sim 32 \%$, respectively. Finally, between 5 and 10~Myr, the median of collisions between embryos increase up 
to $\sim 40 \%$ in the IDA20 simulations, while the median remained the same for the CN08 simulations. For the
collisions with the central object, IDA20 simulations have a median of less than 
$10 \%$, while CN08 simulations median value increased up to $\sim 40 \%$.

After this analysis, we point out that in a standard disk model, the prescriptions
from CN08 and IDA20 lead to differences within the dynamics of the systems that produces
different inner surviving embryo populations not only at the age of disk dissipation, but throughout the entire gas stage.

\subsubsection{Resonances}

During the gas stage, all simulations from both prescriptions show mean motion resonances in the inner population ($a < 0.1~\textrm{au}$). The difference 
between the two treatments lies in the duration and rupture of these resonances.

Two planets whose semimajor axis $a_\textit{i}$ and $a_\textit{j}$ satisfy $a_\textit{i} < a_\textit{j}$, are in commensurable orbits if they follow 
the relation
\begin{equation}
    a_\textrm{i}\approx\left(\frac{q}{p}\right)^{2/3} a_\textrm{j},
\end{equation}
where $p$ and $q$ give the order of the commensurability $p:q$.

\begin{figure}
    \centering
    \includegraphics[width=9cm,height=7cm]{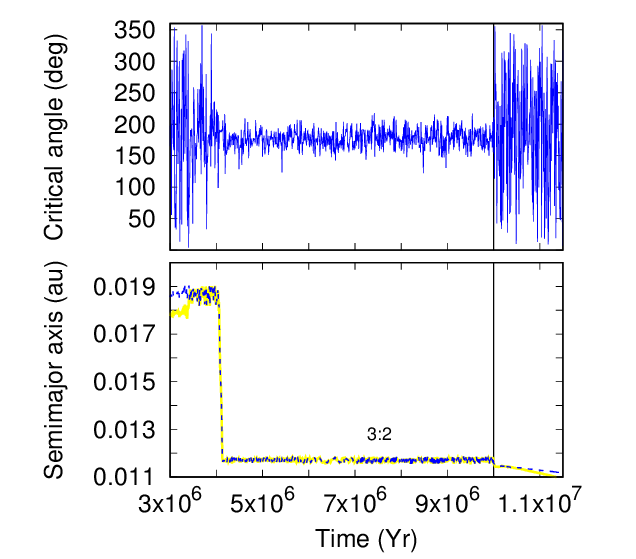}
    \caption{Example of one mean motion resonance under the IDA20 treatment. The critical angle libration (top panel) and commensurability 3:2 (bottom panel) of a pair of planets in resonance are shown during the integration time. The vertical lines represent $t=10~\textrm{Myr}$.}
    \label{fig:ej_resonancia}
\end{figure}

On the other hand, if a pair of planets satisfies this relation and also 
presents a libration critical angle, then this pair of planets is in mean motion resonance. In all our simulations, the critical angles that librated were

\begin{equation}
\theta_1=p\overline{\lambda_\textrm{i}} - q\overline{\lambda_\textrm{j}} - (p-q)\overline{\omega_\textrm{i}},
\end{equation}

\begin{equation}
\theta_2=p\overline{\lambda_\textrm{i}} - q\overline{\lambda_\textrm{j}} - (p-q)\overline{\omega_\textrm{j}},
\end{equation}
where $\overline{\lambda_\textrm{i,j}}=\Omega_\textrm{i,j} + \omega_\textrm{i,j} + M_\textrm{i,j}$ and $\overline{\omega_\textrm{i,j}}=\Omega_\textrm{i,j} + \omega_\textrm{i,j}$, with $\Omega_\text{i,j}$ the node longitude, $\omega_\textrm{i,j}$ the argument of periastron and $M_\textrm{i,j}$ the mean anomaly of each body involved. In Fig. \ref{fig:ej_resonancia} we show one example of a pair of planets under the IDA20 treatment in mean motion resonance, which fulfilled the requirements of a libration angle and being in commensurable orbit in an interval of integration time between $t \sim 4~\textrm{Myr}$ 
and $t \sim 10~\textrm{Myr}$. This pair of planets in mean motion resonance was part of a resonant chain. The pair migrated inward at $t \sim 4~Myr$ after the outer planets involved in the resonant chain collided and caused the remaining innermost planets to migrate inward. When the gas dissipated at $t \sim 10~Myr$, the bodies migrated toward the star at different times, and the mean motion resonance broke apart.

\begin{figure}
    \centering
    \includegraphics[width=9cm,height=6cm]{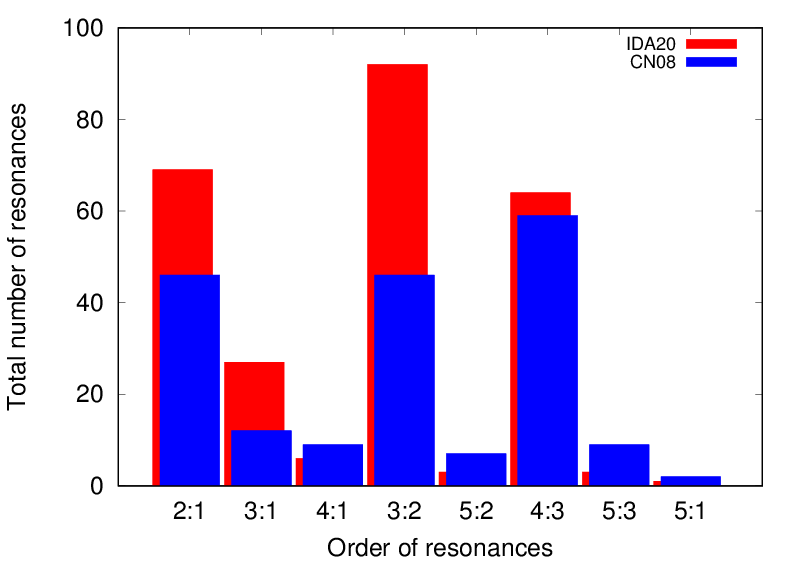}
    \caption{Histogram with the total number of resonances of different orders taking into account the total number of simulations made for each prescription, IDA20 or CN08. An offset was added to the IDA20 resonances for clarity.}
    \label{fig:histo-resonancias}
\end{figure}

For both treatments, the most common resonances are on the order of 2:1, 3:2, and 4:3. However, resonances of order 3:1, 4:1, 5:1, 5:2, and 5:3 were found with lower frequency. We show in Fig. \ref{fig:histo-resonancias} the number of resonances of different orders that we found taking into account all the simulations made under the two prescriptions IDA20 and CN08. We found resonant chains of resonances with different orders throughout the simulations.

\begin{figure*}
    \centering
    \includegraphics[width=18cm,height=6.5cm]{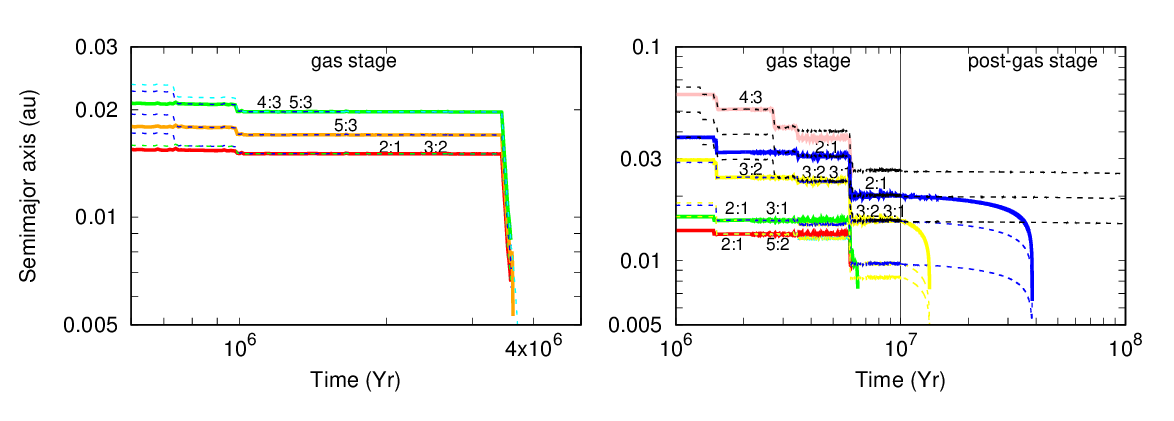}
    \caption{
    Examples of planets in commensurable orbits that represent typical resonant chains from a simulation under the CN08 treatment (left) and the IDA20 treatment (right).}
    \label{fig:resonancias_CN08IDA20}
\end{figure*}

Planets under the CN08 treatment entered in their resonance during the first 
$1~\textrm{Myr}$ and usually remained there for  $1.5-2~\textrm{Myr}$. When the innermost embryo entered the cavity, all the embryos started to move inward, maintaining their respective resonances. The fast migration was mainly caused by the strong gas-disk torques at early stages at distances close to the inner edge of the disk, and second by tidal effects when the embryo was getting closer to the star. The resonances broke apart when the embryos collided with the star. However, we cannot confirm that such a collision caused the breaking as the inward migration occurred in a short period of time ($\sim$40.000 Yr). Then, at the age of $t=5~\textrm{Myr}$, no resonant chain can be found in any of the systems because all the inner bodies 
have already migrated inward and collided with the central object. One example of this is shown in the left panel of Fig. \ref{fig:resonancias_CN08IDA20}, which shows different pairs of planets in a resonant chain that migrated inward after the innermost body entered the cavity. They migrated inward, maintaining their resonances, and finally collided with the star. In most of the cases, no resonant chain was formed later due to a lack of inner planets ($a < 0.1~\textrm{au}$). In less than $\sim 10\%$ of the simulations, one pair of resonant planets could be found from $8~\textrm{Myr}$ to $10~\textrm{Myr}$, which also finally broke up after the gas dissipated from the disk.\\
For planets under the IDA20 treatment, we can distinguish different moments associated with the 
formation of resonant pairs. In all simulations, some pair of planets entered in their resonance during the 
first $1~\textrm{Myr}$. Some of them finally broke apart because one or more of the embryos involved entered the cavity or because they experienced collisions with the embryos involved in the resonant chain, while others remained in resonance until the gas dissipated. On the other hand, some other resonant chains were formed after
$4~\textrm{Myr}$ of integration time, which usually remained during the entire gas stage. One example of this behavior is shown in the right panel of Fig. \ref{fig:resonancias_CN08IDA20}, which shows different pairs of planets in a resonant chain that entered in their resonance at different times. Some resonances broke apart when one pair of embryos collided and others because 
the innermost body entered the cavity. Thousands of years after the innermost planet started to migrate inward, the resonances broke apart and the planets involved started to migrate inward. Then, right after $t\sim10$ Myr, when the gas dissipated, all the surviving resonant chains broke up because the innermost body entered the cavity and started to migrate inward, breaking the other surviving resonances. The migration was mainly caused by the total gas-disk torques and second, by tidal effects when the planet approached the star. In this case, the planets experienced a slower migration than in the CN08 example (in which the bodies entered the 
cavity at $t\sim4~Myr$), as at $t=10~Myr$ the strength of the negative gas-disk torques is lower. Even though no resonant chain was formed after the gas-disk dissipated, some pairs of surviving planets remained close to commensurabilities throughout the integration time.

\subsection{Post-gas stage}
\label{sec:post-gas}

\begin{figure*}
    \centering
    \includegraphics[width=18cm,height=11cm]{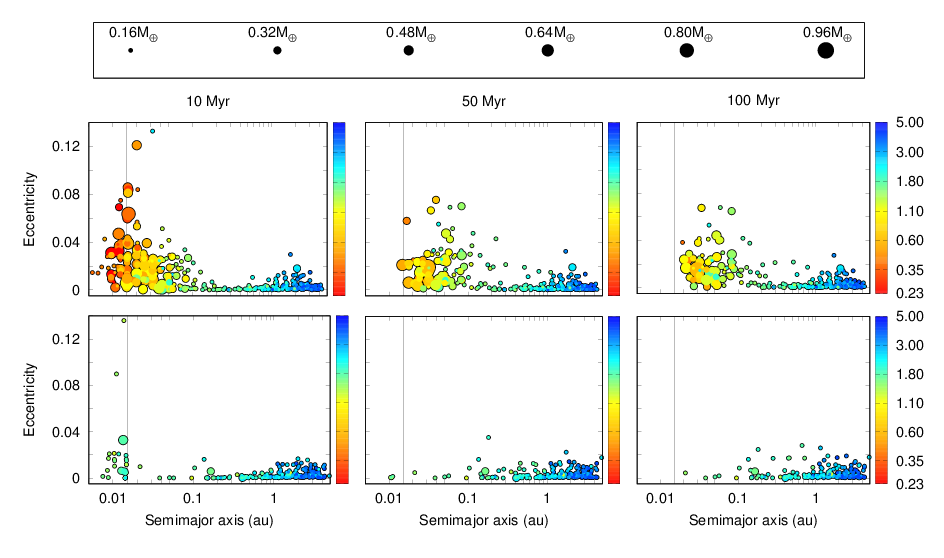}
    \caption{Semimajor axis vs.
    eccentricity for the surviving embryos at ages of $10~\textrm{Myr}$,
    $50~\textrm{Myr}$, and $100~\textrm{Myr}$ for the IDA20 and CN08 
    prescriptions (top and bottom panels, respectively). The colored scale 
    indicates the initial values of the semimajor axis, and the points sizes represent the embryo mass according to the upper labeled black dots. The vertical line indicates the
    position of the inner radii of the gas-disk.}
    \label{fig:ae_postgas}
\end{figure*}

The main difference in the post-gas evolution in both prescriptions 
occurs at distances in the range $0.015 < a/\textrm{au} < 0.1$. 
There, a large population of embryos remains for the IDA20 
prescription with eccentricities up to $\sim 0.08$, while only a few embryos survive in the CN08 
treatment and present quasi-circular 
orbits. At larger distances
($a > 0.1~\textrm{au}$), the effects of the
gas are less relevant and embryos
retain their original quasi-circular orbits.
These results are summarized in Fig. \ref{fig:ae_postgas}, 
which shows the surviving embryos, distinguished by their initial semimajor axis and mass, in all simulations for the two prescriptions at ages of 
$10~\textrm{Myr}$, $50~\textrm{Myr}$, and 
$100~\textrm{Myr}$.

After the dissipation of the 
gas, no other embryo entered the cavity in simulations under the CN08 prescription, while many of them did in simulations under the IDA20 treatment. All the remaining embryos inside the cavity ($a < 0.015~\textrm{au}$) eventually collided with 
the central object. On the one hand, in the simulations under the CN08 treatment, the embryos that entered the cavity just before the gas-disk dissipation were not part of a resonant chain and migrated inward slowly given the weak intensity of the total torque at the inner edge of the disk and the stellar tide. For the simulations under the IDA20 treatment, their evolution toward the central object was mainly caused by gravitational interactions with outer embryos that were part of a resonant chain and were immersed in the gas-disk just before its dissipation. The evolution was also affected by stellar tides when they approached the star $(a\lesssim0.008~au)$.

As explained in Fig. \ref{fig:medians} in Section \ref{sec:gasstage}, most of the collisions occured before the gas dissipation. After this, 
embryos under the CN08 treatment experienced no collisions among themselves, but with the central object, reaching a median value of almost $50\%$ considering all the simulations. On the other hand, embryos under the IDA20 prescription continued to experience collisions among themselves and with the central object up to 50 Myr, increasing the median value up to $\sim45\%$ and $\sim15\%$, respectively. We note that after $50~\textrm{Myr}$, 
no simulation showed a collision between embryos in either of the two prescriptions, while $\sim 22 \%$ and $\sim 40 \%$ of them 
presented one collision with the central object under the CN08 and IDA20 treatments, respectively, even though the median values are the same as at 50 Myr.

\begin{figure*}
    \centering
    \includegraphics[width=0.4\textwidth]{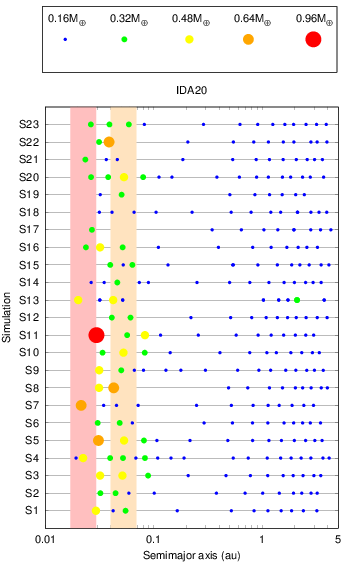}
    \includegraphics[width=0.4\textwidth]{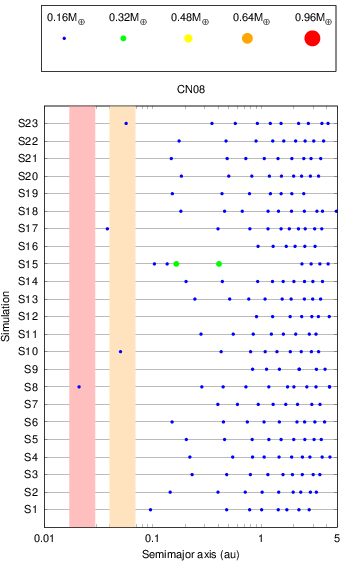}
    \caption{Final planetary architectures for each simulation from the IDA20 prescription (left panel) and CN08 prescription (right panel). The color and size of the points represent the final masses of the planets as shown in
    the upper labeled panels. The cream band represents the IHZ at  $100~\textrm{Myr}$,
    and the pink band shows the IHZ at $1~\textrm{Gyr}$.}
    \label{fig:amasa_postgas}
\end{figure*}

In Fig. \ref{fig:amasa_postgas}, the final architectures of the surviving planets at 100 $\textrm{Myr}$ are 
shown for IDA20 and CN08. Planets are distinguished by size according to their final masses. Most of the CN08 planets remained at their initial mass, while IDA20 
planets present a wide range of masses of up to $\sim 1~M_\oplus$ exclusively in the inner region with $a < 0.1~\textrm{au}$. The isolated habitable zone (IHZ) of the systems at 100 $\textrm{Myr}$ and at 1 
$\textrm{Gyr}$ are also included. We calculated the IHZ according to \citet{Selsis2007} and 
\citet{Barnes2013} and adapted it for our central object of $0.08~M_\odot$ as explained in 
\citet{Sanchez2020}.

Finally, we compared the spatial distribution 
of the resulting embryos with the position of the IHZ. Because of the expected evolution of the 
IHZ toward the central object, we considered its corresponding positions at $100~\textrm{Myr}$ 
and $1~\textrm{Gyr}$ assuming that in this period, the migration and $e$ damping timescales 
due to tidal effects are long enough to avoid a change in the location of the embryos.
The resulting systems from the IDA20 treatment present several embryos inside the IHZ for all the simulations, while from the CN08 prescription only $\sim 10 \%$ of the 
simulations present just one single embryo inside the IHZ. On the one hand, we note that
the embryos inside the $100~\textrm{Myr}$ IHZ will not migrate inward following 
the evolving IHZ. Thus, after few million years, these planets will be located farther away from the external limit of the habitable zone. On the other hand, the embryos located inside the $1~\textrm{Gyr}$ IHZ were unprotected
for a considerable amount of time until the evolving IHZ reached them. The fact that the planets are exposed to stellar radiation inside the inner edge of the IHZ for a long period of time makes the survival of water among other volatiles on the surface of the planets challenging because they experience greenhouse runaway \citep[e.g.,][]{Luger2015}. Thus, the potential habitability of these planets should be carefully evaluated.

\subsection{Comparison with confirmed exoplanets}
\label{sec:observables}

\begin{figure*}
    \centering
   \includegraphics[width=18cm,height=10cm]{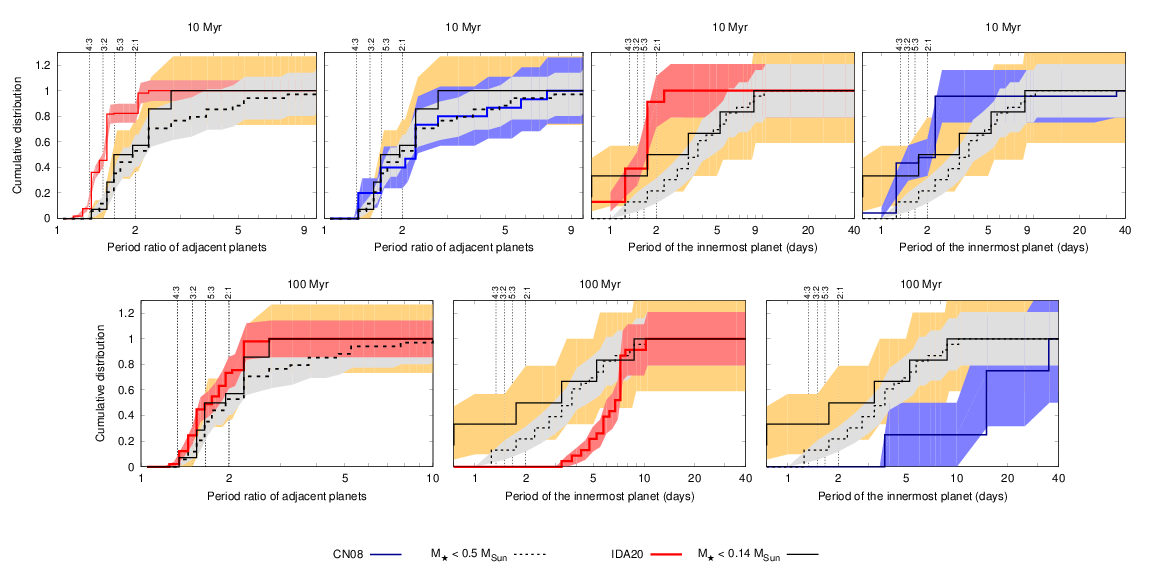}
    \caption{Comparison of cumulative distributions of survival planets in the IDA20 (red line) and CN08 (blue line) simulations at $10~\textrm{Myr}$ (top panels) and at 100 Myr (bottom panels), and the confirmed Earth-like planets around stars with $0.08< M_\star/M_\odot < 0.14~M_\odot$ (solid black line) and $0.14< M_\star/M_\odot < 0.5~M_\odot$ (dotted black line). All planets are located at $a < 0.1~\textrm{au}$. The color shadows correspond to the Poissonian errors in IDA20 (red), CN08 (blue) and confirmed exoplanets around $M_\star < 0.14~M_\odot$ (orange) and $M_\star < 0.5~M_\odot$ (gray).}
    \label{fig:plady_plint}
\end{figure*}

In this section we compare the period distributions of the innermost surviving 
planets ($a < 0.1~\textrm{au}$) with the 
corresponding confirmed Earth-like exoplanets from the Exoplanet 
Archive\footnote{Catalog available at https://exoplanetarchive.ipac.caltech.edu/}. We assumed a mass/radius cutoff of 2 M$_\oplus$/2 R$_\oplus$ and considered the sample of exoplanets detected 
through transit and radial velocity techniques.

As in previous works that studied gas-disk interactions down to the substellar limit \citep[e.g.,][]{Coleman2019,Miguel2019} we compared our results at the end of the gas stage. The top panels of Fig. \ref{fig:plady_plint} show the cumulative
distribution of the period ratio of adjacent planets and
the distribution of periods of the innermost planet 
from each simulation under IDA20 and CN08 at $10~\textrm{Myr}$
together with the corresponding distributions of
confirmed terrestrial exoplanets orbiting stars with masses $0.08<M_\star/M_\odot<0.14$ and  
$0.14<M_\star/~M_\odot<0.5$. We included the Poissonian
errors for the numerical simulations and for 
the exoplanet population around the less and the more massive stars. These errors were calculated with $\pm$ the square of the cumulative planets discretized by period and period ratio, respectively.
The close-in planetary population obtained
following the CN08 prescription around a star of $0.08~M_\odot$ agrees better in the distribution of the period ratio of adjacent planets with the corresponding population from confirmed exoplanets around stars with
masses $M_\star < 0.5~M_\odot$. Despite this similarity, we caution that this comparison was performed against observations of
field stars that are older than $1~\textrm{Gyr}$ and also span a wider mass range than the 
central star of our simulations. On the other hand, IDA20 and CN08 both overestimate the number of planets with innermost periods in comparison with the observations.\\ 

We aim to show how the comparison with the same observations changes at the end of our simulations. The bottom panels of Fig. \ref{fig:plady_plint} show the
distribution of the period ratio of adjacent planets and
the distribution of periods of the innermost planet 
from each IDA20 simulation at $100~\textrm{Myr}$,
together with the corresponding distributions of
confirmed Earth-like exoplanets.
In this case, the close-in planetary population obtained
following the IDA20 prescription shows a distribution of the period ratio of adjacent planets that agrees with the distribution of the confirmed exoplanets around stars with masses $M_\star < 0.14~M_\odot$. A K-S (Kolmogórov-Smirnov) test was performed by applying 500 bootstrap realizations. We found that in only $\sim15\%$ of the realizations we can reject that both distributions came from the same distribution with a 99$\%$ probability. A corresponding analysis is not possible for the simulations
following the CN08 prescriptions because, as we showed in Fig. 
\ref{fig:amasa_postgas} in Section \ref{sec:post-gas}, 
not more than one planet at $a < 0.1~\textrm{au}$ survived at the end of all these simulations. We note that considering gravitational, tidal, and general relativistic effects, the planetary configurations are not expected to change significantly
after $100~\textrm{Myr}$. Thus, it is valid to compare the
simulations with the known exoplanet population around old stars.

We were unable to find a good 
agreement within the innermost planet period distributions regarding the surviving planets in
our simulations under either of the prescriptions (CN08 and IDA20) and the confirmed exoplanets around either of the selected samples. On the one hand, $\sim80\%$ of the surviving innermost planets in simulations under the CN08 treatment have periods longer than 10 days, while all the surviving innermost planets in simulations under the IDA20 prescription have periods between 4 days and 10 days. On the other hand, the innermost observed exoplanet samples have periods between 0.4 days and 10 days. Thus, the innermost planet period distribution taken from the surviving planets in simulations under IDA20 remains closer to the distribution taken from either of the two exoplanet samples, even though it presents a deficit of inner planets.

The deficit of inner planets ($a<0.025~au$) in simulations under the IDA20 treatment would not change on a timescale of a billion years. Considering tidal effects, the resulting planets with $a > 0.025~au$ would decrease their semimajor axis in more than $\sim$10 Gyr. One possible explanation for this deficit might lie in the standard disk model. Changing the disk lifetime, the turbulent and accretion $\alpha$, as well as including an evolving inner edge of the gas-disk might allow the survival of more planets located closer to the star. However, we expect that this consideration will
not produce qualitative differences in the agreement
found within the period ratio of adjacent planets for IDA20 planets and the confirmed Earth-like planets because with the incorporation of inner planets in the systems, the systems would remain in compact configurations.\\

We conclude that it is not correct to compare the resulting planetary configurations at the end of the gas stage with the confirmed exoplanet population. Moreover, this comparison should be made considering an analogous mass for the central object of the systems. We therefore ran the simulations for 100 Myr to be able to compare the planetary architectures with the confirmed exoplanet population. This is as a representative time of the final stage of the system because the number of surviving planets remained almost constant during the last 50 Myr, as shown in Fig. \ref{fig:medians}. Furthermore, regarding gravitational interactions and tidal effects, the planetary configuration would not change significantly in 
$1 Gyr$ given the low eccentricity values, the mass range, and the semimajor axis of the resulting planets.

\section{Conclusions and discussions}
\label{sec:conclusiones}

We treated rocky planet formation around an evolving  $0.08~M_\odot$ star by using $100~\textrm{Myr}$ long $N$-body simulations that incorporate tides and general relativistic effects, as well as gas-disk interactions during the first $10~\textrm{Myr}$. The disk was simulated
according a standard disk model, and its 
effect over the embryos was simulated by independently following the torque prescriptions from IDA20 and CN08. 

We found a resulting close-in ($a < 0.1~\textrm{au}$) planet population under IDA20 that 
did not survive under the CN08 treatment at the end of our simulations
($100~\textrm{Myr}$). Moreover, the IDA20 close-in planet population agrees with the period 
ratio of adjacent confirmed Earth-like exoplanets around stars with $M_\star < 0.14~M_\odot$.
Furthermore, we note that to compare the simulated planetary architectures with the confirmed 
exoplanet population, it is important 
to properly constrain the comparison 
within samples in similar stellar mass 
ranges and using simulations whose integration times are long enough to 
be compared accordingly with the 
old observed planetary systems.

We also found a surviving planet population located in the habitable zone of IDA20 systems that is not 
present in simulations under the CN08 treatment. However, this population should be studied carefully, taking the dynamical evolution history of the planets and the location of the evolving IHZ into account. Considering the influence of gravitational, tidal, and general relativistic effects, no significant changes over time are expected in the planetary configurations after $100~\textrm{Myr}$. Thus, our resulting planets are therefore exposed to stellar radiation due to their location inside the inner edge of
the IHZ for $1~\textrm{Gyr}$. Then, the planets located in the IHZ may not satisfy the sufficient conditions to be considered potentially habitable \citep[e.g.,][]{Luger2015}.\\

We obtained several mean motion resonant chains that survived during the entire gas stage only under the IDA20 prescription. After the gas dissipated, all these resonant chains broke apart. However, some pairs of planets remained close to commensurabilities. This allows the existence of compact planetary systems, which agrees with the period ratio of an adjacent close-in confirmed Earth-like population around less massive stars ($M_\star<0.14M_\odot$).\\
\\
{Even though the resonant breaking mechanism is still not well understood, we aim to discretize three different breaking scenarios that we found in our work}:
\begin{itemize}
    \item \textit{Collision with the central object.} The resonant chains broke apart when the embryos collided with the central object. However, given the fast inward migration, we cannot distinguish whether the breaking was due to the collision with the central object or due to the planets entering the cavity. This is the case for all resonant breaking in simulations under the CN08 treatment.
    \item \textit{Collision among the outer embryos.} The resonances broke apart because the outer embryos involved in the mean motion resonance chain experienced collisions with other embryos during the gas-disk lifetime. This was the case for some of the resonant breaking in simulations under the IDA20 treatment.
    \item \textit{Disk dispersal / disk absence.} Either the resonances broke apart when the innermost planet entered the cavity while the outer planets were still immersed in the disk, or after disk dispersal, when the innermost planet involved in a resonance was located inside the cavity or close to the inner edge of the disk and the outer planet was still immersed in the disk just before the gas-disk dissipation. This was the case of the remaining breaking cases in simulations under the IDA20 treatment.
\end{itemize}
We highlight that the mean motion resonance breaking after the gas-disk dissipation was also found by other authors who studied planet formation around Sun-like stars, even when stellar tides were not considered \citep[e.g.,][]{Izidoro2017,Izidoro2021}.\\

We analyzed tidal and general relativistic effects as well as gas-disk interactions. During the gas stage, the gas-disk interactions play a primary role in the dynamical history of the planets. However, tidal and general relativistic effects give a more detailed model of the orbital dynamic of the planets both during the gas stage and after the gas disk dissipated \citep[][]{Sanchez2020}.\\

In our model, we used a set of initial parameters for the standard gas-disk model. We aim to explore different scenarios for disk masses and lifetimes as well as the impact of different values for the accretion and turbulent $\alpha$, as in \citet{Matsumura2021}, onto the resulting planetary configurations in future works. We expect that for different initial conditions, we may reproduce the innermost exoplanet period distributions 
if an innermost planet population survives in our new numerical simulations.\\

 This study allow us to better understand the rocky planet formation at the substellar mass limit. We conclude that within the framework of parameters we explored, it would be more appropriate to use the IDA20 prescriptions to treat the gas-disk interactions to study rocky planet formation at the substellar mass limit.

\begin{acknowledgements}

This research has made use of the NASA Exoplanet Archive, which is operated by the 
California Institute of Technology, under contract with the National Aeronautics 
and Space Administration under the Exoplanet Exploration Program.
MS and GdE acknowledges the partial financial support by Agencia Nacional de Promoción 
Científica y Tecnológica (ANPCyT) by PICT 201-0505, Argentina, and the partial 
financial support by Facultad de Ciencias Astron\'omicas y Geof\'isicas de la Universidad 
Nacional de La Plata, Argentina (FCAGLP-UNLP) and Instituto de Astrof\'isica de La Plata 
(IALP) for extensive use of their computing facilities.
JJD acknowledges funding from the MIA program at Universidad de la Rep\'ublica, Uruguay, 
and is grateful for the hospitality and support of FCAGLP-UNLP where part of this 
research was carried out. We thank also to the referee Brasser R. for his constructive comments.

\end{acknowledgements}

 \bibliographystyle{aa} 
 \bibliography{Referencias} 

\end{document}